\documentclass[twocolumn, pra, superscriptaddress,notitlepage, nofootinbib]{revtex4-2}
\usepackage{graphicx}
\usepackage{dcolumn}
\usepackage{bm}
\usepackage[usenames,dvipsnames]{color}
\usepackage{mathtools}
\usepackage[most]{tcolorbox}
\usepackage{multirow}
\usepackage{gensymb}
\usepackage[normalem]{ulem}
\usepackage{CJK}
\usepackage{comment}
\usepackage[colorlinks, linkcolor=blue,anchorcolor=blue,citecolor=blue,urlcolor=blue]{hyperref}
\usepackage{amssymb}
\usepackage{pifont}
\usepackage{physics}
\usepackage{natbib}

\usepackage{float} 
\makeatletter 
\let\newfloat\newfloat@ltx 
\makeatother

\usepackage{algorithm}
\usepackage{algpseudocode}
\usepackage{caption}
\usepackage{subcaption}
\usepackage{booktabs}
\usepackage{colortbl}

\begin{document}

\title{Expressive variational quantum circuits provide inherent privacy in federated learning}

\author{Niraj Kumar}
\email{niraj.x7.kumar@jpmchase.com}
\affiliation{Global Technology Applied Research, JPMorgan Chase, New York, NY 10017}
\author{Jamie Heredge}
\affiliation{Global Technology Applied Research, JPMorgan Chase, New York, NY 10017}
\affiliation{School of Physics, The University of Melbourne, Parkville, VIC 3010, Australia}
\author{Changhao Li}
\affiliation{Global Technology Applied Research, JPMorgan Chase, New York, NY 10017}
\author{Shaltiel Eloul}
\affiliation{Global Technology Applied Research, JPMorgan Chase, New York, NY 10017}
\author{Shree Hari Sureshbabu}
\affiliation{Global Technology Applied Research, JPMorgan Chase, New York, NY 10017}
\author{Marco Pistoia}
\affiliation{Global Technology Applied Research, JPMorgan Chase, New York, NY 10017}
\date{\today}

\begin{abstract}
Federated learning has emerged as a viable distributed solution to train machine learning models without the actual need to share data with the central aggregator. However, standard neural network-based federated learning models have been shown to be susceptible to data leakage from the gradients shared with the server. In this work, we introduce federated learning with variational quantum circuit model built using expressive encoding maps coupled with overparameterized ans\"atze. We show that expressive maps lead to inherent privacy against gradient inversion attacks, while overparameterization ensures model trainability. Our privacy framework centers on the complexity of solving the system of high-degree multivariate Chebyshev polynomials generated by the gradients of quantum circuit. We present compelling arguments highlighting the inherent difficulty in solving these equations, both in exact and approximate scenarios. Additionally, we delve into machine learning-based attack strategies and establish a direct connection between overparameterization in the original federated learning model and underparameterization in the attack model. Furthermore, we provide numerical scaling arguments showcasing that underparameterization of the expressive map in the attack model leads to the loss landscape being swamped with exponentially many spurious local minima points, thus making it extremely hard to realize a successful attack.  This provides a strong claim, for the first time, that the nature of quantum machine learning models inherently helps prevent data leakage in federated learning. 
\end{abstract}

\maketitle

\section{Introduction}

The recent advances in the field of machine learning, particularly deep learning, have found tremendous success across a wide variety of areas including computer vision, natural language processing, generative modeling, and time-series forecasting among many others \cite{chowdhary2020natural, ruthotto2021introduction, lim2021time}. With the recent explosion of the abundance of data and powerful computing models, every passing year sees increasing success and impact across the above-mentioned fields using machine learning. 

In parallel, quantum computing, and quantum machine learning \cite{biamonte2017quantum} in particular, has emerged as an interdisciplinary field with the potential to significantly impact a wide range of fields. Some initial results, along with theoretical guarantees, have showcased its potential computational advantage over classical machine learning tasks such as in classification, and distribution learning among others \cite{Biamonte2017, Schuld2014, shaydulin2023evidence}. Some notable examples of quantum machine learning algorithms include the Harrow-Hassidim-Lloyd (HHL) for solving linear systems of equations, quantum principal component analysis, quantum generative models, quantum support vector machines, quantum branch and bound, and quantum Monte-Carlo estimation, among others \cite{harrow2009quantum, lloyd1982least, kerenidis2016quantum, chakrabarti2022universal, montanaro2016quantum, kumar2023q}. 

However, numerous machine learning algorithms require substantial amounts of data to construct robust, high-performing models. In practice, data is often fragmented across various organizations or devices due to factors like privacy regulations (e.g., GDPR \cite{albrecht2016gdpr}) or the logistical challenges of centralizing data collection. For instance, medical data from different hospitals is typically sensitive and confined to individual facilities. This limited and potentially biased data pool makes it challenging to develop accurate models for specific tasks, such as disease pattern identification \cite{rieke2020future}. 

The imperative need for data privacy, coupled with concerns related to storage and computational resources, has seen growing interest in the field of distributed learning. One prominent solution is Federated Learning (FL), which addresses this challenge by forming a loose federation of participating devices, referred to as \emph{clients}, all coordinated by a central server \cite{mcmahan2017communication, yang2019federated}. Importantly, each client has a local training dataset which is never uploaded to the central server. Instead, each client computes an update, typically the model cost function gradient, and transmits only this gradient to the global model maintained by the central server, which we call the \emph{FL model}. The role of the server is to act as a central aggregator of the gradients in order to globally update the FL model. FL approaches have been adopted by various service providers and play a pivotal role in privacy-conscious applications where training data is distributed across multiple clients. Its potential applications are diverse, including sentiment analysis among mobile phone users, enhancing autonomous driving systems, predicting health events such as heart rate risks from wearable devices, improving image detection through data contributed by multiple phone and tablet users, and many more \cite{li2020review, rieke2020future, niknam2020federated, kim2019blockchained, xu2021federated}.

Previous studies have shown that sharing gradient information with the central \emph{honest-but-curious} server can potentially expose clients to sensitive data leakage scenarios \cite{zhao2020idlg, huang2021evaluating}. An honest-but-curious attack model is where the server still performs the gradient aggregation, but could also perform a local gradient inversion-based attack to learn the client's data. This compromise in privacy prevents the FL framework from becoming fully trustworthy for distributed machine learning models. While some studies have indicated that it is possible to partially protect against gradient inversion attacks by employing techniques like gradient masking and introducing random noise \cite{mcmahan2017communication, aono2017privacy}, or by maximizing the mixing of gradients \cite{eloul2022enhancing}, these approaches are not completely robust. They may still leak some data information, or add substantial additional computational overhead, while also in certain cases, reducing the predictive capability of the model. 

In this work, we introduce the quantum version of the federated learning framework where we replace the classical neural network-based FL models with variational quantum circuits (VQC) \cite{benedetti2019parameterized}. We specifically consider VQCs which have a high expressive capability, meaning, they comprise a significant number of non-degenerate Fourier frequencies when representing the model output in the Fourier basis as was originally introduced in \cite{vidal2020input, schuld2021effect} and later explored in \cite{shin2023exponential, landman2022classically, herman2023expressivity}. Our particular emphasis lies in constructing a product encoding map that leads to an exponential increase in the number of frequencies per input dimension when representing the model output and its cost function gradients as functions of the input data. Additionally, the trainable ansatz in these VQCs incorporates a sufficiently large number of trainable parameters within the variational ansatz, rendering them overparameterized \cite{Larocca_2023, anschuetz2022}. In particular, we consider the overparameterized hardware efficient ansatz (HEA) where the number of trainable parameters scales exponentially with the number of qubits $N_q$ \cite{anschuetz2022}. Our choice of expressive encoding map coupled with overparameterized ansatz ensures that the FL model would be able to easily fit a large variety of functions during training and does not suffer the problem of spurious local minima, as all the local minima points exponentially concentrate around the global minimum \cite{schuld2021effect, anschuetz2022}.

Our key contribution is to establish that the gradients generated by the considered VQCs result in a system of \emph{multivariate Chebyshev polynomial equations} with an exponential degree (in the number of qubits) when expressed as a function of input i.e. in the input space\footnote{The gradients can alternatively be written as a system of equations generated by the model output which is linear in the feature space i.e., when represented as the function of an input encoded quantum state. We also study the difficulty of learning the input by solving this system of equations in Sec.~\ref{sec:feature_space}.}. Here, the number of equations is determined by the number of gradients shared, i.e., the number of trainable parameters in the model. Thus, in order to invert the gradients and subsequently learn the data, the task amounts to simultaneously solving these systems of high-degree multivariate equations. We initially tackle the inherent challenge associated with analytically solving these equations, both in exact and approximate scenarios. In the case of exact solutions, we employ techniques from computational algebraic geometry, such as Buchberger's algorithm \cite{buchberger1985}, to provide an upper bound on exact equation solving. For approximate solutions, we utilize the Nyquist-Shannon theorem \cite{shannon1949} to show formulating an approximate system of equations is hard. Both these approaches require time and memory resources scaling exponentially in the number of qubits, thus providing significant privacy in our VQC based FL model.

We also investigate the challenge of retrieving client data via the machine learning technique, the most common gradient inversion attack setting in the standard neural network-based FL models \cite{zhao2020idlg, huang2021evaluating, eloul2022enhancing}. Here, the server generates dummy gradients in their attack model and trains the model with a suitable loss function to match the client's gradients in order to learn the input. Our observation, backed by numerical results, reveals that the attack model in this context faces a significant challenge. It is severely underparameterized, yet highly expressive, meaning it incorporates an exponential number of Fourier frequencies. This results in the loss landscape being riddled with exponentially many spurious local minima which are well separated thus resulting in the untrainability of the attack model, preventing it from easily recovering the original input. This provides compelling evidence that the expressive encoding maps associated with quantum machine learning models are well placed to help prevent data leakage attacks in FL models. 

The next sections are structured in the following manner. Sec.~\ref{relatedwork} discusses related work in the FL and quantum FL literature. We introduce the notion of FL in Sec.~\ref{sec: FL} along with the system of equations generated in the fully connected neural network model which is linear in the input space. Sec.~\ref{sec:qfl} introduces quantum federated learning. In Sec.~\ref{sec:expressivity} we discuss the expressivity of VQCs in general and our specific FL model VQC. Sec.~\ref{sec:Privacy} conceptually highlights the privacy of our quantum FL model which is then further supported by numerical results in Sec.~\ref{sec:numerics}. We finally conclude our results in Sec.~\ref{sec:conclusion}.

\section{Related Work} \label{relatedwork}
Privacy in federated learning models has been challenged by the wider community~\cite{mothukuri2021survey}.  Led by~\cite{Zhu19} and followed by many others~\cite{zhao2020idlg,GeipingBD020,Yin21,ZhuB21}, it was shown that it is possible to extract input data from model gradients. Typical works on inversion attacks in federated learning of classical neural networks share the same principal of guessing the input that generate similar gradients shared by a client. Hence, the attacker goal is to optimize a proxy model to the gradients shared. The results of this attack have shown that it is possible, with relatively ease, to extract data at a pixel-level for full batch images with clarity~\cite{Yin21,ren2022grnn,GeipingBD020}. Bayes frameworks and generative model priors to various input data distributions have been shown to improve inversion success rates, by providing better prior distributions to circumvent several defenses and allow better convergences of various optimizers~\cite{balunovic2021bayesian}. For example,~\cite{zhang2020exploiting} show how to train a generative model prior, resulting in a high success rate of attacks over several image databases. 
Overall, mechanisms to enhance privacy have been proposed, but mainly relying on empirical approaches of noise addition and information compression with a range of effectiveness~\cite{mothukuri2021survey,qammar2022federated,huang2021evaluating, Tramer21}. Such defenses includes dropout nodes, gradient pruning, noise injection to the gradients~\cite{Zhu19,wei2020framework,Tramer21} blending methods of training images applied online by the client nodes~\cite{Huang20}, and also mixing gradients 
strategies~\cite{eloul2022enhancing}. Here, instead of standard machine learning models, we introduce trainable quantum circuits and quantum encoding as a potential resistance strategy for defense. Recent works have shown the applicability of quantum federated learning setup~\cite{chen2021} and potential added value for privacy~\cite{Huang2022,Weikang2021} in using quantum machine learning, however, such intuition was not analyzed nor quantified thoroughly. Here we look into the leakage of data in quantum federated learning using variational quantum circuits, and analyze privacy against a multitude of attack strategies on the client's gradients generated using variational quantum circuits architectures.

\section{Federated Learning} \label{sec: FL}

\subsection{Standard setup} \label{sec:setup}

\begin{figure}[t]
\includegraphics[scale=0.48]{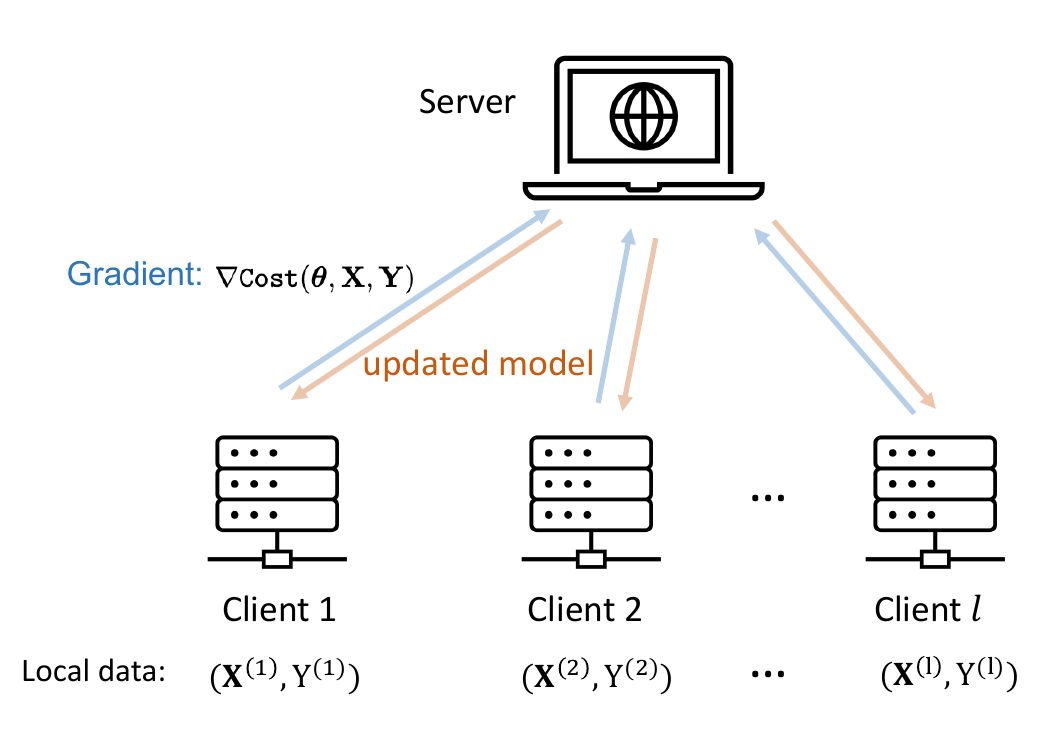}
\caption{Schematic of the federated learning setup with $l$ clients and a central server to perform gradient aggregation. At $t$-th iteration, the server sends the FL model parameters $\boldsymbol{\theta}^t \in \mathbb{R}^d$ to the clients, which then individually generate the cost function gradients of the FL model and send it to the server. The cost function of the client $i \in [l]$ takes as input the model parameters $\boldsymbol{\theta}_t$  and the client's private data $\mathbf{X}^{(i)}, Y^{(i)}$.}
\label{Fig:FL}
\end{figure}

The canonical federated learning problem, as shown in Fig.~\ref{Fig:FL}, involves learning a single, global statistical model from the data stored on multiple remote clients. Consider the setup of $l$ clients with each client $i \in [l]$  having $N_i$ samples of the form, 
$$\mathbf{X}^{(i)}, Y^{(i)} : \{(\mathbf{x}^{(i)}_j, y^{(i)}_j)\}_{j=1}^{N_i}, \hspace{2mm} j \in [l]$$
such that the total number of samples across all the clients is $N = \sum_{i\in [l]} N_i$. Here each input $\mathbf{x}_j^{(i)} \in [0,1]^n$, where $[0,1]^n$ defines all the points within a unit hypercube of dimension $n$, and $y_j^{(i)} \in \mathcal{C}$ for some finite set of output classes where $\mathcal{C} := \{c_1, \cdots, c_{|\mathcal{C}|}\}$.

The aim is to learn the model under the constraint that the client data is processed and stored locally, with only the intermediate updates being communicated periodically with a central server. In particular, the goal is typically to minimize a central objective cost function,
\begin{equation}
   \text{min}_{\boldsymbol{\theta}} \left[ \texttt{Cost}(\boldsymbol{\theta}) = \sum_{i=1}^{l} p_i \texttt{Cost}_i(\boldsymbol{\theta}) \right]
\end{equation}

where $\boldsymbol{\theta} = \{\theta_1, \cdots, \theta_d\} \in \mathbb{R}^d$ are the set of $d$ trainable parameters of the FL model. The user-defined term $p_i \geq 0$ determines the relative impact of each client in the global minimization procedure with two natural settings considered being $p_i = \frac{1}{N}$ or $p_i = \frac{N_i}{N}$. In the setting where the each client $i \in [l]$ performs a batch training with samples $\text{Samp} := [(\mathbf{x}_j^{(i)}, y_j^{(i)})_{j \in \text{Samp}}]$, such that the mini-batch size is $B = |\text{Samp}|$\footnote{We enforce that each client uses the same batch size which is decided apriori.}, then their local cost function $\texttt{Cost}_i(\boldsymbol{\theta})$ is defined as the empirical risk over their local data $f_i(.)$ i.e. 
\begin{equation}
    \texttt{Cost}_i(\boldsymbol{\theta}) = \frac{1}{B} \sum_{j \in \text{Samp}} f_{j}(\boldsymbol{\theta}; (\mathbf{x}_{j}^{(i)}, y_{j}^{(i)}))
\end{equation}
When $B = 1$, this implies that each client randomly picks one sample from their set to compute the cost function. Alternatively, the maximum mini-batch size the clients can pick is $B = \text{min}_i N_i, \hspace{1mm} i \in [l]$.

Note: In the context of classical neural networks, the trainable parameters are the weights and biases of the network $\boldsymbol{\theta}:= \{\mathbf{w}, \mathbf{b}\}$, whereas in the context of variational quantum circuits (as we will see later), the trainable parameters $\boldsymbol{\theta}$ correspond to the tuning of ansatz parameters in the variational quantum circuits. 

\subsection{Task of clients and server}

In the standard federated learning setup, at the $t$-th iteration, the clients each receive the parameter values $\boldsymbol{\theta}^t \in \mathbb{R}^d$ from the server and their task is to compute the gradients with respect to $\boldsymbol{\theta}^t$ and send it back to the server. Here the superscript denotes the iteration step. Upon performing a single batch training, they compute the $d$ gradient updates and share it with the server,
\begin{equation}
     C_{i,j} = \frac{\partial \texttt{Cost}_i(\boldsymbol{\theta}^t)}{\partial \theta^t_j}, \hspace{2mm} \forall i \in [l], j \in [d]
\end{equation}

The server's task is then to perform the gradient aggregation to update the next set of parameters $\boldsymbol{\theta}^{t+1}$ using the rule,
\begin{align}
    \theta_j^{t+1} \rightarrow \theta_j^{t} - \eta \sum_{i=1}^l p_i  C_{i,j}, \hspace{2mm} \forall j \in [d]
\end{align}
where for the rest of the work, we assume the relative impact $p_i = \frac{N_i}{N}$, and $\eta$ is some suitably chosen learning rate hyperparameter by the server. The parameters $\boldsymbol{\theta}^{t+1}$ are then communicated back to the clients and the protocol repeats until a desired stopping criteria is reached.  

\subsection{Data leakage from gradients in the standard Neural Network based FL} \label{sec:fcnn} 

\begin{figure}[t]
\includegraphics[scale=0.35]{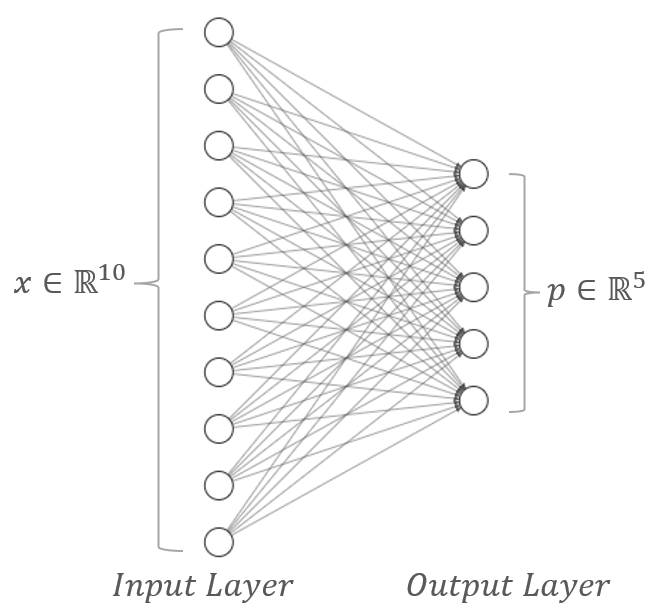}
\centering
\caption{Schematic of fully connected layer where the input $x = x_1x_2\cdots x_{10} \in \mathbb{R}^{10}$ and the output nodes correspond to five output classes.  }
\label{fig:fcnn}
\end{figure}

Here we analyze one major data leakage issue with the standard classical federated learning approach, namely its susceptibility to the gradient inversion attack by the \emph{honest-but-curious} server in successfully recovering the data from the received gradients \cite{eloul2022enhancing, zhao2020idlg}. Let us look at the vulnerability of recovering averaged data information where the model considered for training is the fully connected neural network layer as highlighted in Fig.~\ref{fig:fcnn}. To simplify the analysis, we consider a dense linear layer containing $\mathbf{x} = x_1 \cdots x_n$ as input and $y \in \mathcal{C}$ as output\footnote{We represent the $y$ as $y_1\cdots y_{|\mathcal{C}|}$ such that $y = c_i$ implies $y_i = 1$ and rest being zero. This gives us $|\mathcal{C}|$ number of finite output classes. As an example, for $|\mathcal{C}| = 3$, the set $\mathcal{C} = \{a_1 : 100, a_2 : 010, a_3 : 001\}$.}, where the dense layer is $o_j = \sum_{i=1}^n w_{ij}x_i + b_j$. 
Given a known $y$, $\mathbf{x}$ can be inverted successfully, and any additional hidden layers can be inverted by back-propagation. 

Consider a typical classification architecture which uses softmax activation function, $p_k = \frac{e^{o_k}}{\sum_j e^{o_j}}$ to create the model label $y(\mathbf{w}, \mathbf{b}) = c_{\text{arg max}(p_k)}$, followed by the cross entropy to obtain the cost value,
\begin{equation}
    \texttt{Cost}(p, y) = -\sum_{k}^{\mathcal{C}} y_k \log p_k
\end{equation}
where $\mathcal{C}$ is the number of output classes. The derivative of $p_k$ with respect to each $o_j$ is,
\begin{equation}
\frac{\partial p_k}{\partial o_j} = 
\begin{cases}
& p_k(1 - p_j), \hspace{2mm} k=j \\ 
 & -p_kp_j, \hspace{2mm} k\neq j 
\end{cases}
\end{equation}

Now we can calculate the derivative of the cost function with respect to the weights and biases (via backpropagation),
\begin{equation}
    \frac{\partial \texttt{Cost}}{\partial w_{i, j=k}} = \frac{\partial \texttt{Cost}}{\partial p_k}\frac{\partial p_k}{\partial o_j}\frac{\partial o_j}{\partial w_{i,j}} = (p_j - y_j)x_i
       \label{eqn:weight-grad}
\end{equation}
and,
\begin{equation}
    \frac{\partial \texttt{Cost}}{\partial b_{j=k}} =p_j - y_j
          \label{eqn:bias-grad}
\end{equation}

From this, it can be seen that the number of gradient equations shared with the server would be $n|\mathcal{C}| + |\mathcal{C}|$ while the number of unknowns is $n + |\mathcal{C}|$. For example, from the above equations, it can be seen see that $x_i$ can be found from any $j$, using,
\begin{equation}
   \frac{\partial \texttt{Cost}}{\partial w_{i,j=k}}/\frac{\partial \texttt{Cost}}{\partial b_{j=k}} = x_i 
\end{equation}
In the above setting we saw that for batch size $B=1$, the number of unknowns is less than the number of equations and thus the unknowns can be trivially recovered from the system of equations generated by the dense linear layer of the neural network. 

Let us now see what happens if we consider the mini-batch size training with $B >1$. Here, the client trains with the inputs $\text{Samp} := [(\mathbf{x}_\kappa, y_\kappa)_{\kappa \in \text{Samp}}]$ and only shares the averaged gradient information (over the data points with $B = |\text{Samp}|$) with the server,
\begin{equation}
        \frac{\partial \texttt{Cost}}{\partial w_{i, j=k}} = \frac{1}{B} \sum_{\kappa \in \text{Samp}}(p_{\kappa j} - y_{\kappa j})x_{\kappa i}
\end{equation}
and,
\begin{equation}
    \frac{\partial \texttt{Cost}}{\partial b_{j=k}} = \frac{1}{B} \sum_{\kappa \in \text{Samp}}p_{\kappa j} - y_{\kappa j}
\end{equation}
In this scenario, the number of equations shared is still $n|\mathcal{C}| + |\mathcal{C}|$, whereas the number of unknowns is now $B(n + |\mathcal{C}|)$. Thus the number of unknowns can now exceed the number of equations, resulting in no unique solution for the server when attempting to solve the system of equations. Even in the case that a unique solution exists, numerical optimization can be challenging. However, in the scenario where softmax follows the cross entropy, the authors in \cite{eloul2022enhancing} show that one can obtain an accurate direct solution in many cases even when $B \gg 1$, due to the demixing property across the batch making it easy for the server to retrieve the data points in \text{Samp}.

Certain strategies have been employed in the classical world to prevent the data leakage from the gradients, for example using convolution neural network layers instead of the fully connected layers, or using differential privacy (adding random noise perturbations to the gradients before communicating them with the server), or using classical homomorphic encryption. However, these results either do not guarantee complete data leakage prevention, or prevent data leakage at the cost of reduced model performance or increased communication resource overhead. 

This begs a natural question, 

\noindent \emph{Do quantum machine learning models, when employed in the standard federated learning setup, naturally offer data leakage prevention which the existing neural-network based classical models fail to do?}  

\section{Quantum Federated Learning} \label{sec:qfl}

Quantum federated learning (QFL) has the same setup as the classical federated learning except that the model to be trained is a variational quantum circuit (VQC) instead of a classical neural network. Below we briefly describe the key components of a variational quantum circuit \cite{benedetti2019parameterized} necessary for our construction. Subsequently, we explicitly describe the setup of QFL. 

\subsection{Variational quantum circuit} \label{sec:VQC_circuit}

\begin{figure*}[t]
\includegraphics[scale=0.6]{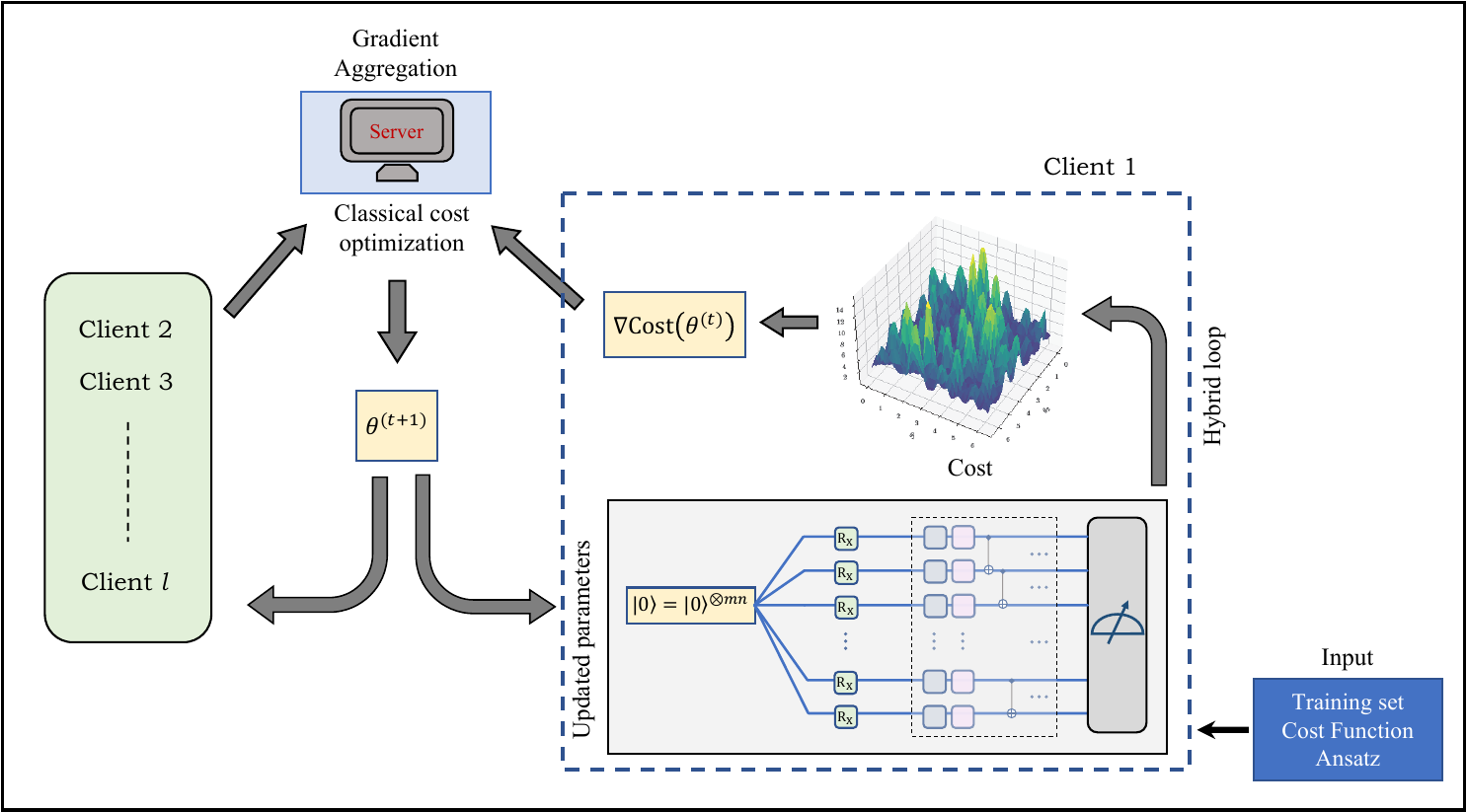}
\caption{ Schematic of the VQC for federated learning: Each client has a VQC circuit consisting of $N_q = nm$ qubits with the encoding map $U(\mathbf{x}) = \bigotimes_{j=1}^n \bigotimes_{k=1}^m R_{X}\left(5^{k-1}\gamma x_{j}\right)$ to load the $n$ dimensional input $\mathbf{x} = x_1\cdots x_n \in [0,1]^n$, where $\gamma = 2\pi$. The encoding map follows the trainable ansatz, in our instance the hardware efficient ansatz consisting of $d$ trainable parameters referred to as $\boldsymbol{\theta}^t \in [0, 2\pi]^d$, where $t$ refers to the iteration step. This is followed by the measurement operator which, in our case is the Pauli operator $\bigotimes_{j=1}^n \bigotimes_{j=1}^m Z$ to produce the model output $y(\mathbf{x}, \boldsymbol{\theta}^t)$. The model output is fed into a chosen cost function $\texttt{Cost}(\mathbf{x}, \boldsymbol{\theta}^t)$ to compute the cost function gradients $\nabla \texttt{Cost}$ with respect to all the $d$ trainable parameters. For this work, we choose the cost function as mean squared error loss. The gradients are then communicated with the server who performs aggregation of the gradients received from all the clients. The server subsequently updates the model and returns updated parameters $\boldsymbol{\theta}^{t+1}$ to all clients.}
\label{Fig:vqc}
\end{figure*}

VQCs are hybrid quantum-classical algorithms where a quantum circuit with trainable parameters is trained iteratively with respect to a classical optimizer as highlighted in Fig.~\ref{Fig:vqc}. The key components of the VQC are: \emph{data encoding map, trainable ansatz, cost function, and gradient estimation}. 

\subsubsection{Unitary to load the classical data} \label{sec:encoding}

Loading the classical input data $\mathbf{x}$ into the quantum circuit is the key to the performance of the VQC. This mapping is defined as,
\begin{equation}
    \mathbf{x} \rightarrow U(\mathbf{\gamma x})
    \label{eq:amp_enc}
\end{equation}

where $\mathbf{x} \in [0,1]^n$, $\gamma$ is the appropriate scaling factor when mapping the classical input into the quantum circuit, and the map $U(.)$ is the unitary map defined on $m$ qubits, where $m$ can in general be different to $n$. For example, in the case of amplitude encoding, the encoding map is,
\begin{equation}
U(\mathbf{x}) = \frac{1}{\|\mathbf{x}\|_2}\sum_{i=1}^n x_i \ket{i}   
\end{equation}
where $m = \mathcal{O}(\log n)$ and the state can be prepared in time $\mathcal{O}(\emph{poly} \log n)$ time using appropriate quantum random access memory model \cite{giovannetti2008quantum, kerenidis2016quantum}.

The choice of the data-uploading unitary is the key to the performance of the VQC as it inherently decides the types of functions (or input/output relationships) that can be learned. In other words, following the results of \cite{schuld2021effect}, the output of a variational quantum circuit can be written in the Fourier function basis where the frequency spectrum is solely determined by the expressive capability of the quantum encoding map. This implies that the VQC model's capability to learn any unknown function increases as one increases the expressivity of the encoding map. In the following sections, we also show that expressivity is directly tied to data privacy in the quantum federated learning setup where the intuition is that a highly expressible encoding map would create a very complex polynomial system of equations for the server to solve which directly ties to the hardness in retrieving the client's data. 

Consequently, in this work, we will consider highly expressive the product feature maps of the type which gives rise to an \emph{exponential encoding scheme} meaning exponential (in the number of qubits) Fourier spectra as we showcase in Sec.~\ref{sec:expressivity}. Specifically the map we consider is the product map, which we refer to as the \emph{Fourier tower map}, considered in \cite{shin2023exponential}. 

The Fourier tower map takes the input $\mathbf{x} \in [0,1]^n$ and realizes the following unitary map on $N_q = nm$ qubits,
\begin{equation}
    U(\mathbf{x}) = \bigotimes_{j=1}^n \bigotimes_{k=1}^m R_{\sigma
}\left(\gamma\beta_{jk}\gamma x_{j}\right)
\label{eq:uploadingmap}
\end{equation}
where the unitary acts on the input state $\ket{0}^{\otimes N_q}$ and $R_{\sigma
}\left(\beta_{jk}x_{j}\right) = \exp\left(-i\gamma\beta_{jk}x_{j}\sigma/2\right)$ is the Pauli rotation operation for the Pauli matrices $\sigma = X, Y,$ or $Z$. Here $m$ is the number of encoding gates chosen per variable of the input $\mathbf{x}$, $\gamma = 2\pi$ is the scaling factor chosen\footnote{The scaling factor of $\gamma = 2\pi$ reflects the fact that the input $x$ lies within a unit hypercube. Thus multiplying the input by $2\pi$ allows the gate $R_{\sigma}(\beta_{jk}x_{j})$ to explore all possible angles from $[0, 2\pi]$ for all $\beta_{jk}\geq 1$.}, and $\beta_{jk}$ is the term added to increase the Fourier spectra of the model as we show in Sec.~\ref{sec:express_our}. In \cite{shin2023exponential}, the authors choose $\sigma = Z$ and showcase that if $\beta_{jk} = 3^{k-1}$, then the model output can be written in a \emph{dense encoding} fashion with a total of $3^m$ frequencies in the range of $[-(3^m - 1)/2, (3^m - 1)/2]$. Here dense encoding implies that the difference between the $k+1$-th and $k$-th frequency is always 1 indicating that this gives rise to a compact encoding. The degree of the Fourier expansion corresponding to the model output $d_F$ is the largest Fourier frequency component. Thus if we want to create a model output with a desired degree $d_F$, then in the case of above encoding, $m = \log_3(2d_F + 1)$. In the context of quantum federated learning, we are interested in the having a compact encoding corresponding to cost function gradient representation in terms of the Fourier series. As we show in Sec.~\ref{sec:express_our}, the corresponding encoding map which gives us the dense encoding is,
\begin{equation}
    U(\mathbf{x}) = \bigotimes_{j=1}^n \bigotimes_{k=1}^m R_{X
}\left(5^{k-1}x_{j}\right)
\label{eq:uploadingmap_2}
\end{equation}

We note that expanding the model's output in terms of Fourier spectra is not the only method to quantify expressivity. Other works have instead considered data uploading maps which gives rise to Chebyshev polynomials of first and second degree \cite{goto2021universal, perez2020data} as we also leverage in showcasing the privacy of quantum federated learning models in Sec.~\ref{sec:privacy}. As Chebyshev polynomials are universal set of overcomplete basis functions, the degree of polynomials generated by the data uploading maps then relates to the expressivity of such quantum models \cite{kyriienko2021solving}.

\subsubsection{Trainable ans\"atze}

Another important aspect of a VQC is its trainable variational ansatz. Generically speaking the form of the ansatz dictates how the parameters $\boldsymbol \theta \in [0, 2\pi]^d$ can be trained in order to minimize the desired cost function.  Without the loss of generality, the ansatz can be expressed as,
\begin{equation}
    W(\boldsymbol \theta) = W_d(\theta_d)W_{d-1}(\theta_{d-1})\cdots W_1(\theta_1)
\end{equation}

where the unitary has the form $W_l(\theta_l) = \exp(-i \theta_l H_l)$, where $H_l$ is the Hermitian operator which generates the quantum gate and $\theta_l$ is the trainable parameter. 

The specific structure of an ansatz will generally depend on the task at hand, as in many cases one can use information about the problem to tailor an ansatz. These are the so-called \emph{problem-inspired ans\"atze} with a few examples being the unitary coupled clustered ansatz, quantum alternating operator ansatz, variable structure ansatz \cite{taube2006new, mcclean2016theory, hadfield2019quantum, farhi2001quantum}. However, some ansatz architectures are generic and \emph{problem-agnostic}, meaning that they can be used even when no relevant information is readily available, for example the hardware efficient ansatz (HEA) \cite{moll2018quantum}. 

In this work we restrict our numerical investigation to HEA although any other choice of the ansatz would be equally valid for quantum federated learning. The structure of HEA we consider, consists of concatenated layers of single qubit rotations of $R_Y-R_Z$ sequenced in each qubit in $N_q$, which are parameterized by independent angles $\theta$ which can produce arbitrary single qubit rotations in $SU(2)$. This is followed by a non-parameterized entangling layer CNOT. The CNOT operation applies between the nearest neighbour qubits. The combination of single qubit layers and CNOT form one block of the ansatz. We apply $L$ such blocks to form the complete ansatz structure such that the total number of trainable parameters in the ansatz is $d = 2LN_q$. The ansatz is then written as,
\begin{equation}
    \small
    W(\boldsymbol{\theta}) = \bigotimes_{i=1}^L\left[\bigotimes_{j=1}^{N_q} R_{Y,j}(\theta_{j,2i-1})R_{Z,j}(\theta_{j,2i})CNOT(j, j+1)\right]
\end{equation}

Combining the data uploading unitary, and the trainable ansatz, the combined unitary can be expressed as,
\begin{equation}
    U(\mathbf{x}; \boldsymbol \theta) = W(\boldsymbol \theta)U(\gamma\mathbf{x})
\end{equation}

\subsubsection{Cost function}

Similar to classical machine learning, for the VQC, one defines the cost function $\texttt{Cost}$ which, in the case of supervised learning with training dataset $\mathbf{X},Y = \{\mathbf{x}_i, y_i\}, i \in N$ where $\mathbf{x}_i \in [0,1]]^n, y_i \in \mathcal{C}$ (see Sec.~\ref{sec:setup}), acts as proxy for the quality of the solution $y(\mathbf{x}_i, \boldsymbol{\theta})$ generated by the model in contrast to the true solution $y_i$. In general, the cost function corresponding to the training dataset can be expressed as,
\begin{equation}
    \texttt{Cost}(\boldsymbol \theta, \mathbf{X}, Y) = \{f_i(y_i, y(\mathbf{x}_i, \boldsymbol{\theta}))\}_{i \in \text{Samp}}
    \label{eq:costgeneral}
\end{equation}
where $f_i$ is some chosen function (for example cross-entropy loss function, mean squared loss function etc.), and $\text{Samp} \subseteq N$. Subsequently, the set of $d$ parameters $\boldsymbol \theta \in [0, 2\pi]^d$ are optimized to train the model and thus solve the optimization task,
\begin{equation}
    \boldsymbol \theta ^* = \text{arg min}_{\boldsymbol \theta} \texttt{Cost}(\boldsymbol\theta)
\end{equation}

In this work, we consider the mean squared error loss as the cost function for train the model,
\begin{equation}
      \texttt{Cost}(\boldsymbol{\theta}, y(\mathbf{x}_i, \boldsymbol{\theta}), y_i) = \left(y_
i -y(\mathbf{x}_i, \boldsymbol{\theta})\right)^2 
      \label{Eq:cost_mse}
\end{equation}

For simplicity, let us denote the output of the VQC model $y(\boldsymbol{\theta}, \mathbf{x}_i)$ by $y_i(\boldsymbol{\theta})$. This can be expressed as,
\begin{equation}
    y_i(\boldsymbol{\theta}) = \text{Tr}[\mathbf{O}W(\boldsymbol{\theta})U(\mathbf{\gamma x_i})\rho_{init}U^{\dagger}(\gamma\mathbf{x}_i)W^{\dagger}(\boldsymbol{\theta})]
\end{equation}
where $\rho_{init}$ is the initial quantum state which is typically $|0^{\otimes N_q}\rangle\langle 0^{\otimes N_q}|$, and $\mathbf{O} = \sum_{i \in |\mathcal{C}|} c_i O_i$ is the measurement operator with $\sum_{i}O_i = \mathbb{I}$, where $O_i$ corresponds to the measurement outcome corresponding to the output class $c_i \in \mathcal{C}$. 

The property of VQC is that they use a quantum computer to estimate the cost value (or its gradient), while leveraging the power of classical optimizers to train the parameters $\boldsymbol \theta$. Some of the desirable criteria for the cost function to meet are, it must be `faithful', meaning the minimum of $\texttt{Cost}(\boldsymbol \theta)$ must correspond to the solution of the problem \cite{bravo2019variational}. And secondly, it must be efficient to estimate $\texttt{Cost}(\boldsymbol \theta)$ by doing quantum measurement and potential classical post processing. Also the cost must be operationally meaningful to guide us through the optimization and it also must be trainable meaning one should be able to optimize over the set of parameters.

\subsubsection{Gradient estimation with parameter shift rule}

Unlike the classical neural network, where the gradients of the cost function are estimated via back propagation, the VQC does not directly offer this possibility. This is primarily because back-propagation would destroy the quantum superposition of the state and thus the data would be irretrievably lost. One could instead perform a finite-difference calculation to estimate the gradients, but this can be too error prone to compute when using hardware, due to inherent gate noise in the quantum system. To resolve this, the authors \cite{mitarai2018quantum, schuld2019evaluating} proposed an analytic version of computing quantum circuit gradients using what is referred as the parameter shift rule. 

A common approach to optimize the cost function $\texttt{Cost}(\boldsymbol \theta)$ is via its gradient, i.e., the change of the function with respect to a variation of its parameters $\boldsymbol \theta$ by a common update rule defined by,
\begin{equation}
    \boldsymbol \theta^{(t+1)} = \boldsymbol \theta^{(t)} - \eta \boldsymbol \nabla  \texttt{Cost}(\boldsymbol \theta^{(t)})
\end{equation}
where $\eta$ is the learning rate. 

So our objective is to compute $\boldsymbol \nabla  \texttt{Cost}(\boldsymbol \theta)$ where the gradient of the cost function can be written as,
\begin{equation}
     \nabla\texttt{Cost}(\boldsymbol \theta, \mathbf{X}, Y) = \Big\{\frac{\partial \texttt{Cost}(\boldsymbol{\theta})}{\partial \theta_1}, \cdots, \frac{\partial \texttt{Cost}(\boldsymbol{\theta})}{\partial \theta_d}\Big\}
\end{equation}
with the gradients with respect to the individual parameters, $\theta_j$ being expressed as,
\begin{equation}
    \frac{\partial \texttt{Cost}(\boldsymbol{\theta})}{\partial \theta_j} = \Big\{f_i\left(y_i, \frac{ \partial y_i(\boldsymbol{\theta})}{\partial \theta_j}\right)\Big\}_{i \in \text{Samp}}
    \label{Eq:costgrad}
\end{equation}

As highlighted above, finite difference to compute the gradient would incur additional errors apart from the statistical error (due to the gate noise). We therefore show how to compute the gradient analytically with the macroscopic shifting of the parameters and subsequent computation of the expectations. 

In order to compute the gradient with respect the cost function, we need to compute the gradient of the model output i.e. $\partial y_i(\boldsymbol{\theta})/\partial \theta_j$. For this, let us express the ansatz $W(\boldsymbol \theta) = UW(\theta_i)V$, were $U, V$ are unitary maps containing other sets of parameters and $W(\theta_j) = \exp(-i \theta_j H_j)$ is parameterized by $\theta_j$ with $H_j$ being the generator for $W$. If $H_j$ has a spectrum of two eigenvalues (which is usually assumed to be the case) labelled by $(\lambda_0, \lambda_1)$, the gradient can be calculated by measuring the observable at two shifted parameter values as follows,
\begin{equation}
    \frac{\partial y_i(\boldsymbol{\theta})}{\partial \theta_j} = r (y_i(\boldsymbol \theta^+) - y_i(\boldsymbol \theta^-))
\end{equation}
where $r = \frac{\lambda_1 - \lambda_0}{2}$ and $\boldsymbol\theta^{\pm} = \boldsymbol \theta \pm \frac{\pi}{4r}\mathbf{e}_j$, where $\mathbf{e}_j$ is a vector with all elements equal to 0 except a single 1 located only at the index $j$. Recent results have also extended this rule to the case where the generator $H_i$ does not satisfy the eigenspectrum condition \cite{kyriienko2021generalized} but for this work, we consider the case where the generator of the ansatz unitary has a two eigenvalue spectrum with $r = 1/2$, as is the case for most Pauli based and control rotation based generators. Thus the model output gradient takes the form, 
\begin{equation}
    \frac{\partial y_i(\boldsymbol{\theta})}{\partial \theta_j} = \frac{1}{2} (y_i(\boldsymbol \theta^+) - y_i(\boldsymbol \theta^-))
    \label{Eq:modeloutgrad}
\end{equation}

In the context of the cost function being mean square error loss function as in Eq.~\ref{Eq:cost_mse}, we can see that the gradient with respect to $\theta_i$ (when cost is defined for single training sample $(\mathbf{x}_i, y_i)$) can be expressed as,
\begin{equation}
    \begin{split}
            \frac{\partial \texttt{Cost}(\boldsymbol \theta)}{\partial \theta_j} &= -2 \left(y_i -y_i(\boldsymbol \theta)\right) \frac{\partial y_i(\boldsymbol \theta)}{\partial \theta_j} \\
    &= -\left(y_i -y_i(\boldsymbol \theta)\right) (y_i(\boldsymbol \theta^+) - y_i(\boldsymbol \theta^-))
    \label{Eq:cost_ms_grad}
    \end{split}
\end{equation}

The above cost gradient estimation is with respect to training the model with a single sample at a time i.e. batch training with batch size $B = 1$. If on the other hand, the cost function is estimated for $\text{Samp} := (\mathbf{X}, Y) = [(\mathbf{x}_i, y_i)_{i \in \text{Samp}}]$, with batch size $B = |\text{Samp}|$, then the gradient is expressed as,
\begin{equation}
    \begin{split}
        \frac{\partial \texttt{Cost}(\boldsymbol \theta)}{\partial \theta_j} &= -\frac{2}{B}\sum_{i \in \text{Samp}}\left(y_i -y_i(\boldsymbol \theta)\right) \frac{\partial y_i(\boldsymbol \theta)}{\partial \theta_j} \\
    &= -\frac{1}{B}\sum_{i \in \text{Samp}}\left(y_i -y_i(\boldsymbol \theta)\right) \left(y_i(\boldsymbol \theta^+) - y_i(\boldsymbol \theta^-)\right)
    \label{Eq:cost batch_ms_grad}        
    \end{split}
\end{equation}

\subsection{Federated learning with VQC} \label{sec:FLVQC}

We briefly revisit the federated learning setup defined in Sec.~\ref{sec: FL} in the context of VQC. The setup consists of $l$ clients with each client $i \in [l]$  having $N_i$ samples of the form, 
$$\mathbf{X}^{(i)}, Y^{(i)}: \{(\mathbf{x}^{(i)}_j, y^{(i)}_j)\}_{j=1}^{N_i}, \hspace{2mm} i \in [l]$$ such that the total number of samples across all the clients is $N = \sum_{i\in [l]} N_i$. 
The aim is to learn the underlying relationship between the example vectors and their labels by training the VQC model under the constraint that the client data is processed and stored locally, with only the intermediate updates being communicated periodically with a central server. In particular, the goal is to minimize the mean squared error loss function as in Eq.~\ref{Eq:cost_mse} with the gradient-based method.

In the setup, at the $t$-th iteration, the clients each receive the parameter values $\boldsymbol{\theta}^t$ from the server and their task is to compute the gradients with respect to $\boldsymbol{\theta}^t$ and send it back to the server. In the case of batch training with batch size $B = 1$, the clients compute their gradients according to Eq.~\ref{Eq:cost_ms_grad}, or alternatively with Eq.~\ref{Eq:cost batch_ms_grad} when $B > 1$.  Upon performing a single batch training, each client $i$ shares the $d$ gradient updates with the server referred as,
\begin{equation}
     C_{i,j} = \frac{\partial \texttt{Cost}_i(\boldsymbol{\theta}^t)}{\partial \theta^t_j}, \hspace{2mm} \forall i \in [l], j \in [d]
     \label{eq:grad_share}
\end{equation}

The server's task is then to perform the gradient aggregation to update the next set of parameters $\boldsymbol{\theta}^{t+1}$ using the rule,
\begin{align}
    \theta_j^{t+1} \rightarrow \theta_j^{t} - \alpha \sum_{i=1}^l p_i  C_{i,j}, \hspace{2mm} \forall j \in [d]
\end{align}

where  $p_i = \frac{N_i}{N}$, and $\alpha$ is some suitably chosen learning rate hyperparameter by the server.

Thus it becomes clear that in a single batch training, each client shares $d$ gradient updates (and hence $d$ equations) with the server, while the number of unknowns from the server's perspective is $B(n + |\mathcal{C}|)$ for each client (given that the inputs $\mathbf{x}_i \in \mathbb{R}^n$ and the target label can be in any of one of the $|\mathcal{C}|$ classes). 

Next, we consider the question of \emph{privacy} of client's data from the \emph{honest-but-curious} server's perspective where the task of the server effectively boils down to extracting the unknowns from the shared gradient
information. Here we consider the \emph{worst-case} data-leakage scenario for the clients where we quantify the hardness of learning even a single input-output data from the clients i.e. batch size $B=1$. In order to prove privacy, we connect privacy with expressivity of VQCs. In the next section, we quantify the expressivity of the quantum model in terms of the frequency spectra generated by the VQC as shown by \cite{schuld2021effect}.

\section{Expressivity of VQC} \label{sec:expressivity}

Here we first showcase that a general VQC model output and the cost function gradient can be expressed as the sum of Fourier basis functions. Subsequently, we quantify the expressivity of our VQC model (as defined in Sec.~\ref{sec:VQC_circuit}) in terms of non-degenerate Fourier spectra. 

\subsection{General VQC Fourier expansion} \label{sec:gen_fou}

\subsubsection{Univariate input analysis}

Let us start our analysis with the univariate input $x \in [0,1]$ and characterize the function $y(\boldsymbol{\theta})$. The univariate function that the quantum circuit outputs with respect to an observable $\mathbf{O}$ is then,
\begin{equation}
\small
    y(\boldsymbol{\theta}) = \text{Tr}[\mathbf{O} W(\boldsymbol\theta)U(\gamma x)|0^{\otimes N_q}\rangle\langle 0^{\otimes N_q}|U^{\dagger}(\gamma x)]W^{\dagger}(\boldsymbol \theta)]
\end{equation}
where $\boldsymbol{\theta} := \{\theta_1, \cdots, \theta_d\} \in [0, 2\pi]^d$ and the circuit is defined on $N_q = m\times 1 = m$ qubits.

Now, an encoding unitary can always be written as $U(x) = e^{-i\gamma xH}$ for some generator Hamiltonian $H$. Furthermore, one can always write the eigenvalue decomposition of the Hamiltonian as $H = V^{\dagger}\Sigma V$, where $\Sigma$ is the diagonal matrix with eigenvalue entries $\lambda_1,\cdots, \lambda_p$. The encoding unitary becomes  $U(\gamma x) = V^{\dagger}e^{-i\gamma x\Sigma} V$. Let us assume the entries of the matrix V are $[V]_{ij} = v_{ij}$, and the entries of ansatz are $[W(\boldsymbol \theta)]_{ij} = w_{ij}$ where $i, j \in [1, 2^{N_q}]$ (since we are defining the circuit of $N_q$ qubits). 

Let us first compute the quantum state after the application of the encoding map and the trainable ansatz, $\ket{\psi} = W(\boldsymbol \theta)U(\gamma x)\ket{0}^{\otimes N_q} = W(\boldsymbol \theta) V^{\dagger} e^{-i\gamma x \Sigma} V \ket{0}^{\otimes N_q}$. One can verify that the state takes the form,
\begin{equation}
    [\ket{\psi}]_{j} = \sum_{k=1}^p \sum_{l=1}^p e^{-i \gamma x\lambda_k} w_{jl} v_{kl}^{*} v_{k1} \hspace{2mm} j\in [1, p] 
\end{equation}
where $\ket{\psi}$ can be written as a column vector with $2^{N_q}$ entries of which the first $p$ entries are non zero with entries given as in the above equation while the rest $2^{N_q} - p$ entries are zero. 

Similarly, the $j$-th entry for the row vector $\bra{\psi}$ is,
\begin{equation}
        [\bra{\psi}]_{j} = \sum_{k=1}^p \sum_{l=1}^p e^{i\gamma x\lambda_k} w_{jl}^* v_{kl} v_{k1}^* \hspace{2mm} j\in [1, p] 
\end{equation}

Next we want to express $y(\boldsymbol\theta)$,
\begin{equation}
    \begin{split}
            y(\boldsymbol\theta) &= \bra{\psi}\mathbf{O} \ket{\psi} \\
    &= \sum_{a=1}^p \sum_{b=1}^p \mathbf{O}_{ab} [\bra{\psi}]_a [\ket{\psi}]_b \\
    &= \sum_{k=1}^p \sum_{l=1}^p e^{i\gamma x(\lambda_k - \lambda_l)} \left( \sum_{a=1}^p \sum_{b=1}^p \mathbf{O}_{ab} c^*_{ak} c_{bl}\right) \\
    &= \sum_{\omega \in \Omega} A_\omega e^{i\gamma x\omega}
    \label{eq:for_gen_uni}
    \end{split}
\end{equation}

where in the second step, $$c_{ij} = \sum_{k=1}^p w_{ik}v^*_{jk}v_{j1}$$
Also in the last step of Eq.~\ref{eq:for_gen_uni}, we group the frequencies $\omega = \lambda_k - \lambda_l$ for $k, l \in [p]$. Thus all the frequencies accessible to the model are,
\begin{equation}
    \Omega = \{\lambda_k - \lambda_l, \hspace{2mm} k,l \in [p]\}
\end{equation}
and the coefficients $$A_\omega = \sum_{\lambda_k - \lambda_l = \omega}\sum_{a,b=1}^ d O_{ab} c^*_{ak} c_{bl}$$

Few things to note here are that $0 \in \Omega$, and if $\omega \in \Omega$, then $-\omega \in \Omega$. Also we have that $A_\omega = A^*_{-\omega}$, thus $y(\boldsymbol \theta)$ represents a real valued function. Consequently, the size of the spectrum of $y(\boldsymbol{\theta})$ is K, 
\begin{equation}
    K = \frac{|\Omega| - 1}{2}
\end{equation}
where in order to compute the size of the spectrum, we exclude the $0$ frequency and count $\omega$ and $-\omega$ as a single frequency. The degree of Eq.~\ref{eq:for_gen_uni} is defined as the largest available frequency $D = \text{max}(\Omega)$.

Once we have the model output $y(\boldsymbol{\theta})$ expressed in the Fourier basis function, the cost function and the gradients can also be expressed in the same basis. Considering the mean squared error as the cost function for input sample $(x,y)$, its expansion in the Fourier basis is the following,
\begin{equation} \label{eq:cost_univ}
    \begin{split}
          \texttt{Cost} &= \left(y -y(\boldsymbol \theta)\right)^2 \\
      &= y^2 + y(\boldsymbol{\theta})^2 - 2yy(\boldsymbol{\theta}) \\
      &= y^2 + \sum_{\omega, \omega' \in \Omega} A_{\omega} A_{\omega'}e^{i\gamma x(\omega + \omega')} - \sum_{\omega \in \Omega} 2y A_{\omega}e^{i\gamma x\omega} \\
      &= \sum_{\omega \in \Omega} \mathcal{A}_{\omega,\omega'}e^{i\gamma x(\omega + \omega')} \\
    \end{split}
\end{equation}

where in the last expression the coefficient $\mathcal{A}_{\omega,\omega'}$ encapsulates the three separate terms in the third term of the equation. Further, we define the set $\Omega_g$ such that,
\begin{equation}
\begin{split}
    \Omega_g &= \{\omega_g = \omega + \omega', \hspace{2mm} \omega, \omega' \in \Omega\} \\
    &= \{(\lambda_k - \lambda_l + \lambda_y - \lambda_z, \hspace{2mm} k,l,y,z \in [p]\}
\end{split}
\end{equation}

Now we can express Eq.~\ref{eq:cost_univ} as,
\begin{equation}
    \texttt{Cost} = \sum_{\omega_g \in \Omega_g} \mathcal{A}_{\omega_g}e^{i\gamma x\omega_g}
\end{equation}

where the coefficient of the different frequency terms $\omega_g$ are,
\begin{equation}
    \mathcal{A}_{\omega_g} = \sum_{\underset{\omega + \omega' = \omega_g}{\omega, \omega' \in \Omega}} \mathcal{A}_{\omega, \omega'}
\end{equation}

Taking the gradient of the cost expression with respect to the parameters $\boldsymbol{\theta} \in \{\theta_1, \cdots, \theta_d\}$ results in,
\begin{equation}
    \begin{split}
        C_1 &= \frac{\partial}{\partial \theta_1} \texttt{Cost}(y( \boldsymbol\theta), y) = \sum_{\omega_g\in \Omega_g} \frac{\partial \mathcal{A}_{\omega_g}}{\partial \theta_1}e^{i\gamma x\omega_g}  \\
    & \hspace{35mm}\vdots \\
    C_d &= \frac{\partial}{\partial \theta_d} \texttt{Cost}(y(\boldsymbol\theta), y) = \sum_{\omega_g \in \Omega_g} \frac{\partial \mathcal{A}_{\omega_g}}{\partial \theta_d}e^{i\gamma x\omega_g}        
    \end{split}
    \label{eqn:gradients_equations_univariate}
\end{equation}
We note that the size of the spectrum for the gradients $\{C_1,\cdots, C_d\}$ is larger than the size of the spectrum for $y(\boldsymbol{\theta})$ because the available frequencies enter as $\omega_g = \omega + \omega'$, instead of $\omega$. We denote the size of the set $\Omega_g$ by $|\Omega_g|$. The total size of the spectrum is (excluding 0 frequency and counting $\omega_g$ and $-\omega_g$ as one frequency),
\begin{equation}
    K_g = \frac{|\Omega_g| - 1}{2}
\end{equation}
Another point to note here is that while the frequency spectrum $\Omega_g$ decides the number of frequencies that the model gradient function can generate and consequently decides the maximum degree of polynomials that the model gradient can represent, it is also the flexibility of the coefficients $\mathcal{A}_{\omega_g}$ which plays a key role in final expressivity of the model gradients functions $\{C_1,\cdots, C_d\}$. The expressivity of model coefficients is ultimately controlled by the parameters $\boldsymbol{\theta}$  as the Fourier coefficients are not all arbitrary but rather constrained by the degrees of freedom/expressivity of the ansatz. Thus the VQC model with $d$ real trainable parameters has the capability of arbitrarily controlling $d/2$ complex coefficients $\mathcal{A}_{\omega_g}$ out of a total $K_g$ coefficients \cite{schuld2021effect}. 

From the perspective of showing privacy of quantum federated learning as introduced in Sec.~\ref{sec:Privacy}, we care about the coefficients (especially for high degree frequencies) being non-zero. From the numerical results of \cite{landman2022classically}, it turns out the that the Fourier Tower feature map (as defined in Eq.~\ref{eq:uploadingmap_2}) combined with HEA ends up having non-zero high degree Fourier coefficients (even though they are lower in magnitudes compared to the frequencies that exhibit a high number of redundancies).

\subsubsection{Multivariate input analysis}

The previous section was the stepping stone to the more relevant case in quantum federated learning where the input $\mathbf{x} \in [0,1]^n$ is loaded into the quantum circuit via the unitary $U(\mathbf{\gamma x})$. The quantum model output is then of the form,
\begin{equation}
    \small
    y(\boldsymbol{\theta}) = \text{Tr}[\mathbf{O} W(\boldsymbol\theta)U(\gamma \mathbf{x})|0^{\otimes N_q}\rangle\langle 0^{\otimes N_q}|U^{\dagger}(\gamma \mathbf{x})]W^{\dagger}(\boldsymbol \theta)]
\end{equation}
where $\boldsymbol{\theta} := \{\theta_1, \cdots, \theta_d\} \in [0, 2\pi]^d$. Similar to the previous analysis, the encoding unitary can be expressed as,
\begin{equation}
    U(\gamma\mathbf{x}) = e^{-i\gamma(x_1H_1 + x_2H_2 + \cdots + x_nH_n)} 
\end{equation}
where in this work, we assume the data encoding unitary to be a \emph{product map} meaning all the Hamiltonians commute i.e. $[H_i, H_j] = 0, i,j \in [n]$. Thus the above unitary can be expressed as,
\begin{equation}
    U(\mathbf{x}) = e^{-i\gamma x_1H_1} \otimes \cdots \otimes e^{-i\gamma x_n H_n}
\end{equation}
Again, after the eigenvalue decomposition of the Hamiltonians $H_i, i \in [n]$, let the resulting eigenvector matrix be $V_i$ and the eigenvalues be $\lambda_1^{(i)}, \cdots, \lambda^{(i)}_p$. Thus the encoding unitary has the simultaneous eigenvalue decomposition of the form,
\begin{equation}
U(\gamma \mathbf{x}) = V^{\dagger}S(\gamma \mathbf{x})V 
\end{equation}
where $S(\gamma \mathbf{x}) = \left(\bigotimes_{j=1}^n e^{-i\gamma x_j\Sigma_j}\right)$, $V = V_1\otimes\cdots \otimes V_n$, and $\Sigma_j$ is the diagonal matrix with entries $\lambda_1^{(j)},\cdots, \lambda^{(j)}_{2^m}$, where $\lambda^{(j)}_{k} = 0, k \in [p+1, 2^m]$. Here $U(\mathbf{x})$ and thus $V$ are block-diagonal matrices defined on $N_ q = nm$ qubits with $V_i, i\in [n]$ being defined on $m$ qubits. 

Let us denote $\boldsymbol{\lambda}_{\mathbf{k}} := (\lambda_{k_1}^{(1)},\cdots, \lambda^{(n)}_{k_n})$, where the index $\mathbf{k} := k_1\cdots k_n$ with each $k_i \in [p]$ (given that for each Hamiltonian $H_i$, the number of distinct eigenvalues  are $p$). Using this bold-index notation, we can write the diagonal elements of $S(\gamma \mathbf{x})$ as,
\begin{equation}
    [S(\gamma \mathbf{x})]_{\mathbf{k},\mathbf{k}} = e^{-i \gamma\mathbf{x}\cdot \boldsymbol{\lambda}_{\mathbf{k}}}
\end{equation}

where $\mathbf{x}\cdot\boldsymbol{\lambda}_{\mathbf{k}} = \sum_{i=1}^n x_i \lambda_{k_i}^{(i)}$. 
Further, the elements of the matrices $M := \{V, W(\boldsymbol{\theta}), \mathbf{O}\}$ are also written in this multi-index notation as $[M]_{\mathbf{i,j}} = m_{\mathbf{i,j}}$.

Similar to the previous univariate analysis, we see that $y(\boldsymbol\theta)$ can be expressed as,
\begin{align}
    y(\boldsymbol\theta) &= \bra{\psi}\mathbf{O} \ket{\psi} \\
    &= \sum_{\mathbf{a}, \mathbf{b} \in [p]^n} \mathbf{O}_{\mathbf{a,b}} [\bra{\psi}]_{\mathbf{a}} [\ket{\psi}]_{\mathbf{b}} \\
    &= \sum_{\mathbf{k}, \mathbf{l} \in [p]^n} e^{i\gamma \mathbf{x}\cdot(\boldsymbol{\lambda}_{\mathbf{k}} - \boldsymbol{\lambda}_{\mathbf{l}})} \left( \sum_{\mathbf{a}, \mathbf{b} \in [p]^n} \mathbf{O}_{\mathbf{a,b}} c^*_{\mathbf{a,k}} c_{\mathbf{b,l}}\right) \\
    &= \sum_{\boldsymbol\omega \in \boldsymbol\Omega} A_{\boldsymbol\omega} e^{i\gamma \mathbf{x}\cdot\boldsymbol\omega}
    \label{eq:for_gen_uni_mult}
\end{align}

where we use the multi-index notation $\mathbf{a, b, k, l} \in [p]^n$ to denote the elements of the matrix. Also, in the second step, $$c_{\mathbf{i,j}} = \sum_{\mathbf{k} \in [p]^n} w_{\mathbf{i,k}}v^*_{\mathbf{j,k}}v_{\mathbf{j,1}}$$ Thus, we see that, the frequencies accessible to the model are,
\begin{equation}
    \boldsymbol\Omega = \{\boldsymbol{\omega} := \boldsymbol{\lambda_k - \lambda_l}, \hspace{2mm} \boldsymbol{k,l} \in [p]^n\}
\end{equation}
and the coefficients 
\begin{equation}
A_{\boldsymbol\omega} =  \sum_{\boldsymbol{\lambda_k - \lambda_l} = \boldsymbol{\omega}}\sum_{\mathbf{a}, \mathbf{b} \in [p]^n} \mathbf{O}_{\mathbf{a,b}} c^*_{\mathbf{a,k}} c_{\mathbf{b,l}}
\label{eq:coefficients}
\end{equation}

Let us denote $\boldsymbol{\Omega} = (\Omega_1, \cdots, \Omega_n)$, where $\Omega_j = \{\omega_j = \lambda^{(j)}_{k_j} - \lambda^{(j)}_{l_j}, \hspace{2mm} k_j, l_j \in [p]\}$. From this it becomes clear that the size of $\boldsymbol{\Omega}$ is the just the Cartesian product of the sizes of individual $\Omega_j$ i.e.,
\begin{equation}
    |\boldsymbol\Omega| = \prod_{j=1}^n |\Omega_j| 
\end{equation}

Thus the size of the frequency spectrum $(K)$ in the general multivariate case is,
\begin{equation}
    \begin{split}
        K &= \frac{1}{2}(|\boldsymbol\Omega| - 1) \\
        &= \frac{1}{2}\left(\prod_{j=1}^n |\Omega_j| - 1\right)
    \end{split}
\end{equation}

Similar to the univariate case analysis, the mean squared error cost function for the input sample $(\mathbf{x}, y)$ can be written as,
\begin{equation}
    \begin{split}
          \texttt{Cost} &= \left(y -y(\boldsymbol \theta)\right)^2 \\
      &= \sum_{\boldsymbol{\omega, \omega'} \in \boldsymbol{\Omega}} \mathcal{A}_{\boldsymbol{\omega,\omega'}}e^{i\gamma \mathbf{x}\cdot(\boldsymbol{\omega + \omega'})}   \\
      &= \sum_{\boldsymbol{\omega_g} \in \boldsymbol{\Omega_g}} \mathcal{A}_{\boldsymbol{\omega_g}}e^{i\gamma \mathbf{x}\cdot\boldsymbol{\omega_g}}  \\
    \end{split}
\end{equation}

where the set of accessible frequencies for the cost function are,
\begin{equation}
    \boldsymbol{\Omega_g} = \{\boldsymbol{\omega_g} = \boldsymbol{\omega + \omega'} : (\boldsymbol{\lambda_k}-\boldsymbol{\lambda_l}) + (\boldsymbol{\lambda_y} - \boldsymbol{\lambda_z)}\}
\end{equation}
where $\boldsymbol{k,l,y,z} \in [p]^n$. Taking the gradient of the cost function with respect to the parameters $\boldsymbol{\theta} \in \{\theta_1, \cdots, \theta_d\}$ results in,
\begin{equation}
    C_j = \frac{\partial}{\partial \theta_j} \texttt{Cost}(y( \boldsymbol{\theta}), y) = \sum_{\boldsymbol{\omega}_g \in \boldsymbol{\Omega_g}} \frac{\partial \mathcal{A}_{\boldsymbol{\omega_g}}}{\partial \theta_j}e^{i\gamma \mathbf{x}\cdot\boldsymbol{\omega_g}}   
    \label{eqn:multivariate_gradient_omega}
\end{equation}
where $j \in [d]$. 
Now we can denote the set as $\boldsymbol{\Omega_g} = (\Omega_{1g},\cdots,\Omega_{ng})$, where,  $$\Omega_{jg} = \{\omega_{jg} = (\lambda^{(j)}_{k_j} - \lambda^{(j)}_{l_j}) + (\lambda^{(j)}_{y_j} - \lambda^{(j)}_{z_j})\}$$ with $k_j, l_j, y_j, z_j  \in [p]\}$. 

Thus the size of $\boldsymbol\Omega_g$ is again just the Cartesian product of individual $\Omega_{jg}$ and this the size of the spectrum $K_g$ is,
\begin{equation}
    \begin{split}
        K_g &= \frac{1}{2}(|\boldsymbol\Omega_g| - 1) \\
        &= \frac{1}{2}\left(\prod_{j=1}^n |\Omega_{jg}| - 1\right)
    \end{split}
\end{equation}

\subsection{VQC in this work}

\subsubsection{Encoding map to load classical data} \label{sec:express_our}

As we highlight in Sec.~\ref{sec:gen_fou}, the ability of a quantum model to learn any unknown function (relationship of input-output) is directly tied to to its Fourier spectrum size and the coefficients corresponding to distinct frequencies. Thus, as mentioned in Sec.~\ref{sec:encoding},  we consider a highly expressive and dense product encoding map which gives rise to an exponential encoding scheme, meaning exponential (in the number of qubits per input dimension) Fourier spectra. Specifically the map we consider is the product map, which we refer to as the \emph{Fourier tower map} which takes the input $\mathbf{x} \in [0,1]^n$ and realizes the following unitary map on $N_q = nm$ qubits,
\begin{equation}
    U(\gamma \mathbf{x}) = \bigotimes_{j=1}^n \bigotimes_{r=1}^m R_{X
}\left(5^{r-1}\gamma x_{j}\right)
\end{equation}

where the unitary acts on the input state $\ket{0}^{\otimes N_q}$ and $R_{X
}\left(5^{r-1}\gamma x_{j}\right) = \exp\left(-i5^{r-1}\gamma x_{j}X/2\right)$ is the Pauli $X$ rotation operation, and $\gamma = 2\pi$. The reason to choose the prefactor $5^{r-1}$ to ensure we build a \emph{dense} Fourier spectra of the cost function gradient output where the dense spectra implies that the difference between the $k+1$-th and $k$-th frequency is always 1 indicating that this gives rise to a compact encoding.

Let us first analyse this in the univariate case where the input $x \in [0,1]$ and the encoding is defined on $m$ qubits i.e. in this case $N_q = m$ and the encoding map is written as,
\begin{equation}
    U(\gamma x) = e^{-i\gamma xX/2} \otimes e^{-i5\gamma xX/2} \cdots \otimes e^{-i5^{m-1}\gamma xX/2} 
\end{equation}

The corresponding generator Hamiltonian of the unitary $U(\gamma x)$ can be written as,
\begin{equation}
    H = \frac{1}{2}\left(X\otimes\cdots\mathbb{I} + 5\mathbb{I}\otimes X\otimes\cdots\mathbb{I} + \cdots 5^{m-1}\mathbb{I}\otimes \cdots X\right)
\end{equation}
It can be easily checked that this Hamiltonian has $2^m$ eigenvectors denoted by $|\boldsymbol{\tilde{k}}\rangle = |\tilde{k_1}\cdots \tilde{k_m}\rangle$, where $|\tilde{k_j}\rangle =  \frac{1}{\sqrt{2}}(\ket{0} + (-1)^{k_j} \ket{1}$ with $k_j \in \{0,1\}$, and thus $\boldsymbol{k} = k_1\cdots k_m \in \{0,1\}^m$. The corresponding eigenvalue for $|\boldsymbol{\tilde{k}}\rangle$ is then,
\begin{equation}
    \lambda_{\boldsymbol{k}} = \sum_{j=1}^m (-1)^{k_j}\frac{5^{j -1}}{2}
\end{equation}

We are now interested in quantifying the Fourier spectrum of cost function gradient resulting from the model output. For this, we can write equation Eq.~\ref{eq:cost_univ} as,
\begin{equation} 
    \begin{split}
          \texttt{Cost} &=\sum_{\omega_g \in \Omega_g} \mathcal{A}_{\omega_g}e^{i\gamma x\omega_g} \\
          &= \sum_{\boldsymbol{k,l, y,z} \in \{0,1\}^m} \mathcal{A}_{\boldsymbol{k,l, y,z}}e^{i\gamma x(\lambda_{\boldsymbol{k}} - \lambda_{\boldsymbol{l}} + \lambda_{\boldsymbol{y}} - \lambda_{\boldsymbol{z}})} 
    \end{split}
\end{equation}

Thus the set of accessible frequencies can be written as,
\begin{equation}
    \Omega_g = \{\omega_g = (\lambda_{\boldsymbol{k}} - \lambda_{\boldsymbol{l}}) + (\lambda_{\boldsymbol{y}} - \lambda_{\boldsymbol{z}})\}
\end{equation}
where $\boldsymbol{k,l,y,z} \in \{0,1\}^m$. One can also write the set of accessible frequencies as,
\begin{equation}
\begin{split}
    \Omega_g &= \left\{\sum_{j=1}^m \frac{1}{2}\left((-1)^{k_j} - (-1)^{l_j} + (-1)^{y_j} - (-1)^{z_j}\right)5^{j-1}\right\}\\
    &= \{c_j 5^{j-1}\}
\end{split}
\end{equation}
where $k_j, l_j, y_j, z_j \in \{0,1\}$. An immediate observation is that $c_j = \{-2, -1, 0, 1, 2\}$. Thus $\Omega_g$ would have all possible integer values in the range,
\begin{equation}
\begin{split}
        \Omega_g &= \left[\sum_{j=1}^m -2\cdot5^{j-1}, \sum_{j=1}^m 2\cdot5^{j-1}\right] \\
        &= \left[\frac{-(5^m-1)}{2}, \frac{5^m -1}{2}\right]
\end{split}
\end{equation}
This gives us the total size of the set $\Omega_g$ which is,
\begin{equation}
    |\Omega_g| = 5^m
\end{equation}

Thus the total size of the spectrum in the univariate case is $K_g = (|\Omega_g|-1)/2 = (5^m - 1)/2$. We also see the maximum degree of the Fourier component is $d_F = (5^m - 1)/2$. This implies that in order to realise the cost function gradient output with a certain fixed degree $d_F$, the number of qubits ($m$) needs to be,
\begin{equation}
    m = \log_5(2d_F + 1)
\end{equation}

This can be easily extended to the multivariate case of $\mathbf{x} \in [0,1]^n$ where the size of the set of accessible frequencies $\boldsymbol{\Omega_g}$ is,
\begin{equation}
    |\boldsymbol{\Omega_g}| = |\Omega_g|^n = 5^{mn}
\end{equation}

This gives us the total size of the spectrum as,
\begin{equation}
    K_g = \frac{|\boldsymbol{\Omega_g}| - 1}{2} = \frac{5^{mn}-1}{2}
\end{equation}

\subsubsection{Overparameterized Ans\"atze}

In addition to privacy, we also desire the FL models to be expressive and trainable. For this precise reason, we consider our trainable ansatz to be overparameterized, which we implement in this case using a number of trainable parameters that scale exponentially in the number of qubits \cite{Larocca_2023, anschuetz2022}. Overparameterization offers a multitude of benefits in the FL model, namely this makes the quantum model \emph{fully expressive} by allowing complete control in tuning of parameterized coefficients in Eq.~\ref{eq:coefficients}. Further, as highlighted in \cite{anschuetz2022}, overparameterization substantially improves the trainability of FL model by getting rid of spurious local minima i.e. the local minima are no longer well separated from the global minimum but rather concentrated around the global minimum. This further implies that any trainability issues in retrieving client input during a gradient inversion attack is isolated to the attack side of training and reflects an increased privacy in the model. 

For our trainable ansatz, we consider the hardware efficient overparameterized ansatz where the number of trainable parameters scale exponentially with the number of qubits $N_q = mn$, such that,
\begin{equation}
    \text{Num. param} \geq 4^{N_q}
\end{equation}
This is a sufficient condition for overparameterisation of the hardware efficient ansatz  \cite{anschuetz2022}. 

\section{Privacy in Quantum Federated Learning} \label{sec:Privacy}

Privacy of local data is one of the main highlights of the use of federated learning. As already mentioned in Sec.~\ref{sec:fcnn}, typical classical federated learning setups are prone to data leakage of the clients which is a substantial drawback in the existing setups and something that is addressed quite naturally with the use of expressive variational quantum circuits. 

As highlighted in Sec.~\ref{sec:FLVQC}, privacy is defined in the context where the honest-but-curious server aims to learn the client's data (unknowns) given a set of gradient updates. Furthermore, we consider privacy in the \emph{worst-case} data leakage setting for the clients, where the clients execute their quantum circuits perfectly without any noise and compute and share the cost function gradients exactly. We showcase that even in this ideal setting, the server is unable to easily learn the client's data  We note that this is the best case scenario for the server trying the learn the data, as a setting with experimental errors would introduce further noise in the cost function gradients, thus making it harder to learn the data input \cite{wei2020federated}. Here we showcase the privacy when the cost function chosen by all the clients is the mean-squared error. We note that similar results can be obtained for cross entropy or other chosen cost functions.

\subsection{Information available to the server} \label{sec:info}

Let us start by quantifying the amount of information available to the server from each client,
\begin{itemize}
    \item A total of $d$ gradient information updates $\frac{\partial}{\partial \theta_j}\texttt{Cost}(\boldsymbol \theta)$ of the form of Eq.~\ref{eq:grad_share}.
    \item The batch size $B$ employed by clients to perform the training of FL model. 
    \item Encoding map architecture $U$ as highlighted in Figure~\ref{Fig:vqc} (the input parameters $\mathbf{x}$ is unknown but the server knows the quantum gate that is being applied to encode the input). 
    \item Ansatz architecture $W$ as well as the parameters $\boldsymbol\theta$ (the server can completely replicate this process).
    \item Measurement operator $\mathbf{O}$.
    \item The cost function being applied by the clients (the mean squared error in our case). 
\end{itemize}

\subsection{Privacy: Definition} \label{sec:privacy}

 Consider the setting of Sec.~\ref{sec:FLVQC} where a client performs batch training with batch size $B$. Then for any given batch and any given client, the number of gradient updates shared with the server is $d$ (Eq.~\ref{eq:grad_share}) while the number of unknowns is $B\cdot(n + \mathcal{|C|})$. Thus, for each client, the honest-but-curious server needs to learn $(\mathbf{x}_j, y_j)$, $\forall j \in \text{Samp}, B = |\text{Samp}|$\footnote{For simplicity we denote the mini-batch used by the client $i \in [l]$ to compute the gradients on as $\text{Samp} := [(\mathbf{x}_j, y_j)_{j \in \text{Samp}}]$ with the mini-batch size $B = |\text{Samp}|$.}. 

Here, we study the \emph{stricter version} of privacy of data-leakage to the server by considering the case when the mini-batch size $B=1$ i.e. for each client's input $(\mathbf{x}, y)$, when $d$ gradient information of the form Eq.~\ref{Eq:cost_ms_grad} are shared to server, we study the success probability for the server to output the sample $(\mathbf{x}', y')$  such that,
\begin{equation}
    \|\mathbf{x'} - \mathbf{x} \| \leq \epsilon_1, \hspace{2mm} |y' - y| \leq \epsilon_2
\end{equation}
for any desired $\epsilon_1, \epsilon_2 > 0$.

One thing that immediately becomes obvious when choosing the cost function as mean-squared error loss, is that the gradients are of the form Eq.~\ref{Eq:cost_ms_grad},
\begin{equation}
            \frac{\partial \texttt{Cost}(\boldsymbol \theta)}{\partial \theta_j} 
    = -\left(y -y(\boldsymbol \theta)\right) (y(\boldsymbol \theta^+) - y(\boldsymbol \theta^-))
\end{equation}
If the output label class $y \in \mathcal{C} = \{-1, +1\}$, then it becomes trivial for the server to know the label $y$ from the sign of the gradient. One can also extend this to general multi-class case $\mathcal{C}$ in terms of recovering the true label value $y$. Thus, here we assume the worst case label leakage scenario where the server can ascertain the value of the labels perfectly. In this case the only hidden part for the server is $\mathbf{x}$. It turns out that leaking the label information is usually not an issue in federated learning as the client's data that contains the sensitive information is $\mathbf{x}$ rather than $y$. Thus we recast the objective of the server in this setting to be able to output,
\begin{equation}
    \|\mathbf{x'} - \mathbf{x} \| \leq \epsilon
\end{equation}
 
for some chosen value of $\epsilon$.

\subsection{Attack strategy for the server}

A general attack strategy of the server in order to learn the data input is \emph{to solve a system of multivariate polynomial functions} in the input space i.e. the polynomials are a function of the input $\mathbf{x}$. Let us see how we can rewrite the problem of learning the inputs from the gradient information into the problem of solving system of multivariate equations. 

Consider the problem setting where each client has a variational quantum circuit as shown in Figure~\ref{Fig:vqc}. At the $t$-th iteration step, they receive the parameters $\boldsymbol{\theta}^t  \in [0, 2\pi]^d$ from the server, where the superscript denotes the iteration step. For simplicity, we refer to $\boldsymbol{\theta}^t$ as $\boldsymbol{\theta}$. As highlighted in the previous sections, the clients compute their gradients by considering the mini-batch size $B = 1$, i.e., they each pick a training sample $(\mathbf{x}, y)$ from their local dataset and compute the cost function $\texttt{Cost}(\boldsymbol \theta)$ with the mean squared error loss function as defined in Eq.~\ref{Eq:cost_mse}. From this they can compute the gradients of the cost function according to Eq.~\ref{Eq:cost_ms_grad} which we write as a function of $(\boldsymbol{\theta}, \mathbf{x}, y)$. We denote these gradients as,
\begin{equation}
    C_j = \frac{\partial}{\partial \theta_j}\texttt{Cost}(\boldsymbol \theta, \mathbf{x}, y), \hspace{2mm} \forall j \in [d]
    \label{eq:target-gradient}
\end{equation}
where for simplicity we denote the quantity $C_{i,j}$ in Eq.~\ref{eq:grad_share} as $C_j$.

It becomes clear that upon sharing the gradient values $\{C_1,\cdots, C_d\}$ with the server, the task of the server is to learn $\mathbf{x}'$ (here we assume that they can exacly infer $y$ from the gradients) such that,
\begin{equation}
f_j(\boldsymbol \theta, \mathbf{x}', \mathbf{y}) =  \frac{\partial}{\partial \theta_j}\texttt{Cost}(\boldsymbol \theta, \mathbf{x}', \mathbf{y}) - C_j = 0, \hspace{2mm} \forall j \in [d]
\label{eq:system_of_eq}
\end{equation}
i.e., the model gradient generated by $\mathbf{x}'$ must match the gradients received from the clients.

This is a general system of $d$ multivariate polynomial equations $f_j$ with $n$ unknowns $\mathbf{x}' = x'_1\cdots x'_n$. Here the multivariate polynomial equations emerge upon expanding the derivative of the cost function with respect to the unknown $\mathbf{x}$ as we see in the following sections. We note that this formalism is completely general and thus also encompasses the classical fully connected Neural network setting we saw in section~\ref{sec:fcnn} where the system of equations generated are linear in the inputs and thus becomes easy to invert in the settings where the number of equations exceed the number of unknowns. 

\subsubsection{Generating system of Chebyshev polynomials with quantum circuits}

In the case of fully connected classical neural networks, it is seen in Eq.~\ref{eqn:weight-grad} and \ref{eqn:bias-grad} that the honest-but-curious server's objective of learning the client's data reduces to solving the system of equations which are linear in the inputs. The equivalent situation is much harder in the quantum setting as we will now demonstrate by reformulating the gradients generated from the quantum model as high degree Chebyshev polynomials.

As shown in Eq.~\ref{eqn:gradients_equations_univariate}, the individual gradients which the server generates with their attack model in the univariate case $x' \in [0,1]$ can be written as a Fourier series with the set of accessible frequencies being $\Omega_g$,
\begin{equation}
      C'_j = \frac{\partial}{\partial \theta_j} \texttt{Cost}(\boldsymbol{\theta}, x', y) = \sum_{\omega_g\in \Omega_g} \frac{\partial \mathcal{A}_{\omega_g}}{\partial \theta_j}e^{i\gamma x'\omega_g}  
\end{equation}
where $j \in [d]$. Given that the gradients are real values, this enforces the condition on the coefficients of the Fourier expansion to satisfy,
\begin{equation}
    \frac{\partial \mathcal{A}_{\omega_g}}{\partial \theta_j} = \frac{\partial \mathcal{A}^*_{-\omega_g}}{\partial \theta_j}
\end{equation}
Using this fact, the gradient can be written as,
\begin{equation}
\small
C'_j = \frac{\partial \mathcal{A}_{0}}{\partial \theta_j} +  \sum_{\omega_g =1}^{d_F} 2\left[ \Re (\frac{\partial \mathcal{A}_{\omega_g}}{\partial \theta_j}) \cos{\omega_g \gamma x'} - \Im (\frac{\partial \mathcal{A}_{\omega_g}}{\partial \theta_j}) \sin{\omega_g \gamma x'} \right] 
\end{equation}
where $d_F$ is the maximum degree of the cost function frequency set $\Omega_g$\footnote{Note that this is true given our encoding circuit defined in Sec.~\ref{sec:express_our} where the gradient frequencies are integers and dense i.e. separated by 1.}. 

The Chebyshev polynomials are defined as, 
\begin{equation}
    \begin{split}
    T_{\omega_g}(\cos{\gamma x'}) &\equiv \cos{(\omega_g \gamma x')} \\
    U_{\omega_g-1}(\cos{\gamma x'}) \sin{\gamma x'} &\equiv \sin{(\omega_g \gamma x')}    
    \end{split}
\end{equation}
The gradients can therefore be written in terms of Chebyshev polynomials as,
\begin{equation}
    \begin{split}
        C'_j = \frac{\partial \mathcal{A}_{0}}{\partial \theta_j} &+  \sum_{\omega_g = 1}^{d_F} \Big( 2 \Re (\frac{\partial \mathcal{A}_{\omega_g}}{\partial \theta_j}) T_{\omega_g}(\cos{\gamma x'}) \\
        &- 2 \Im (\frac{\partial \mathcal{A}_{\omega_g}}{\partial \theta_j}) U_{\omega_g-1}(\cos{\gamma x'}) \sin{\gamma x'} \Big)
    \end{split}
\end{equation}

We now can perform a change of variables $c = \cos{\gamma x'}$ and $s = \sin{\gamma x'}$, which introduces an additional constraint $c^2 + s^2 = 1$. If we subtract this expression from the target gradients, we are left with a system of $d + 1$ polynomials in two variables $(c, s)$, that we wish to solve. The maximum degree of this polynomial will be equal to the maximum frequency of the spectrum which is $d_F = \frac{5^m - 1}{2}$ as mentioned in Sec.~\ref{sec:express_our}. Absorbing constants into the target gradient $C_j$ received from the client, the series of polynomials can be written as 

\begin{equation}
\small
    \begin{split}
        C_1 - \sum_{\omega_g = 1}^{d_F}  \Big( 2 \Re (\frac{\partial \mathcal{A}_{\omega_g}}{\partial \theta_1}) T_{\omega_g}(c) - 2 \Im (\frac{\partial \mathcal{A}_{\omega_g}}{\partial \theta_1}) U_{\omega_g-1}(c) s  \Big) = 0 & \\
             \hspace{35mm}\vdots \hspace{40 mm} & \\
        C_d -  \sum_{\omega_g = 1}^{d_F}\Big(  2 \Re (\frac{\partial \mathcal{A}_{\omega_g}}{\partial \theta_d}) T_{\omega_g}(c) - 2 \Im (\frac{\partial \mathcal{A}_{\omega_g}}{\partial \theta_d}) U_{\omega_g-1}(c) s \Big) = 0 &\\      c^2 + s^2 - 1 = 0 &
    \end{split}
\end{equation}

Extending this to the multivariate input case $\mathbf{x'} \in [0,1]^n$, we have that the cost function gradient generated by the server is,
\begin{equation}
\small
\begin{split}
        C'_j &=\frac{\partial}{\partial \theta_j} \texttt{Cost}(\boldsymbol{\theta}, \mathbf{x}', y) = \sum_{\boldsymbol{\omega_g} \in \boldsymbol{\Omega_g}} \frac{\partial \mathcal{A}_{\boldsymbol{\omega_g}}}{\partial \theta_j}e^{i\gamma \mathbf{x}'\cdot\boldsymbol{\omega_g}} \\
        &= \sum_{\boldsymbol{\omega_g} \in \boldsymbol{\Omega_g}} 2\text{Re} ( \frac{\partial  \mathcal{A}_{\boldsymbol{\omega_g}}}{\partial \theta_j}) \cos(\gamma\mathbf{x'}\cdot \boldsymbol{\omega_g})\\
        &\hspace{13mm}- 2 \text{Im} ( \frac{\partial  \mathcal{A}_{\boldsymbol{\omega_g}}}{\partial \theta_j}) \sin(\gamma \mathbf{x'}\cdot \boldsymbol{\omega_g})
        \label{eq:chebyshev}
\end{split}     
\end{equation}
Due to the fact that $\mathbf{x'}\cdot \boldsymbol{\omega_g} = \sum_{k=1}^n x'_k\omega_{gk}$, an explicit form of Eq.~\ref{eq:chebyshev} would require the iterative application of trigonometric angle addition formulas, such as $\cos{(a + b)} = \cos{(a)}\cos{(b)} - \sin{(a)}\sin{(b)}$. We will therefore focus on finding the highest degree of the resulting polynomial rather than writing an explicit form. Tracking only the first term in this iterative expansion we see that one of terms will be, 
\begin{equation}
    \prod_{k=1}^n \cos(\omega_{gk} x_k ) = \prod_{k=1}^n T_{\omega_{gk}}(\cos{x_k}) = \prod_{k=1}^n T_{\omega_{gk}}(c_k)
\end{equation}
The maximum value of any $\omega_{gk}$ is $d_F$ and hence the maximum degree of the multivariate polynomial for $n$ data inputs will be $d_F$. Hence in the general multivariate case, the system of polynomial equation would have $2n$ unknowns $[c_k, s_k]$ with a maximum degree of $  (d_F)^n = (5^m - 1)^n/2^n$. 

\subsubsection{Analytical solution to Chebyshev polynomials}

There are two main challenges to finding a solution to this system of equations that are not present in the classical case previously described. The first is calculation of the coefficient terms $\frac{\partial \mathcal{A}_{\boldsymbol{\omega_g}}}{\partial \theta_j}$, and the second is solving the system of high degree polynomials.

Analytically finding the coefficients would involve symbolically simulating the entire quantum circuit in order the find the gradients of the cost function of the respective circuit model output. This is clearly impractical from a memory and time complexity perspective as the Hilbert space dimension of a quantum state scales exponentially as $2^{nm}$. 

Even if these coefficients are known, then the problem would still involve solving a system of polynomial equations with $2n$ unknowns and max degree $(d_F)^n$. There are techniques in computational algebraic geometry such as Buchberger's algorithm \cite{buchberger1985}, that can be used to find exact solutions. The worst case complexity scaling of Buchberger's algorithm has been shown to scale as,
\begin{equation}
    2\Big( \frac{\Delta^2}{2} + \Delta \Big) ^{2^{N-2}} 
\end{equation}
where $\Delta$ is the maximum total degree of any of the polynomials and $N$ the number of unknown variables \cite{dube1990grobner}. Hence finding an analytical solution utilizing Buchberger's algorithm would be upper bounded by a scaling of  $\mathcal{O}(\sqrt{5}^{mn4^n})$. The combination of these two issues provides significantly more safety against attacks involving analytical solution finding approach to learn the client's input. 

\subsubsection{Approximate solution to system of equations}

Digressing from the pursuit of an exact analytical solution, we can instead explore the feasibility of obtaining approximate numerical solutions for solving the system of Chebyshev polynomial equations. In the context of approximate equation solving, the challenge of determining the coefficients still persists. In the case of solving the system of equations approximately, the task of finding the coefficients still remains. The Nyquist–Shannon sampling theorem offers insights in this regard \cite{shannon1949}. It states that for a function with the highest Fourier frequency $d_F$, at least $2d_F$ samples of the function will be required to avoid aliasing effects i.e., the effect resulting in overlapping frequency components resulting in poor coefficient reconstruction. Therefore, in the univariate case, at least $2d_F = 5^m - 1$ samples of the cost function gradient (Eq.~\ref{eq:chebyshev}) would be required to obtain the coefficients for that gradient. In the multivariate $n$ dimensional case, the number of samples required would be $(2d_F)^n = (5^m - 1)^n$ given that one would have to sample in $n$ dimensions. From here, one needs to obtain the coefficients of the Chebyshev polynomial and subsequently, numerically solve the resulting equations. As discussed previously, this will introduce additional complexity depending on the numerical algorithm utilized. 

Since calculating the coefficients, either exactly or approximately, already scales exponentially in $m$, we will explore alternative numerical methods that do not require the calculation of the coefficients, namely \emph{machine learning-based gradient inversion attacks}. We then argue in further sections that even these attacks require optimization algorithms that will exponentially scale in $m$ and therefore expressive quantum models provide inherent privacy in federated learning.

\subsubsection{Gradient inversion machine learning attack}

As the previous sections indicate, analytical or even approximate solutions to the system of equations are exponentially hard. We therefore focus our numerical results on utilizing the machine learning-based gradient inversion approach. Given all the information available to the server, their attack strategy to learn the data using the machine learning-based minimization technique is given in Algorithm~\ref{alg:attack}. 

\begin{algorithm}
\caption{Gradient inversion machine learning attack}\label{alg:attack}
\begin{algorithmic}
\Require $\{\frac{\partial}{\partial \theta_1}\texttt{Cost}(\boldsymbol \theta),\cdots,\frac{\partial}{\partial \theta_d}\texttt{Cost}(\boldsymbol \theta)\}, W(\boldsymbol \theta), U$
\Ensure Optimize to learn $(\mathbf{x}, y)$. 
\State 1. Construct a VQC such that output state is $\ket{\mathbf{x}', \boldsymbol \theta} = W(\boldsymbol \theta)U(\mathbf{x}')\ket{0}$.
\State 2. Construct the cost gradient $\frac{\partial}{\partial \theta_j}\texttt{Cost}(\boldsymbol \theta, \mathbf{x}', y)$ using the parameter shift rule or other methods.
\State 3. Repeat the procedure for all $j \in [d]$.
\State 4. Minimize the loss function of the attack model, $$L = \sum_{j=1}^d  L_j\left(\frac{\partial}{\partial \theta_j}\texttt{Cost}(\boldsymbol \theta, \mathbf{x}', y) , \frac{\partial}{\partial \theta_j}\texttt{Cost}(\boldsymbol \theta, \mathbf{x}, y)\right)$$ by optimizing over $\mathbf{x}'$, where $L(.)$ is a chosen loss function.
\State 5. Learn $\mathbf{x}'$ such that it is $\epsilon$ close to $\mathbf{x}$ in a distance measure appropriate for the data.
\end{algorithmic}
\end{algorithm}

In terms of attempting an approximate solution using gradient descent, we are required to solve a loss function that we chose to be the $l_2$ loss function, 
 \begin{equation}
     L = \sum_{j=1}^d  \left(\frac{\partial}{\partial \theta_j}\texttt{Cost}(\boldsymbol \theta, \mathbf{x}', y) -  \frac{\partial}{\partial \theta_j}\texttt{Cost}(\boldsymbol \theta, \mathbf{x}, y)\right)^2
 \end{equation}
by optimizing over $\mathbf{x}'$ until the gradients match the target gradients. We will refer to the value optimized by the FL model during training as the cost function $\texttt{Cost}$ and refer to the value optimized during the gradient inversion attack as the loss function $L$ of the attack model.

In order to solve this with a gradient descent-based attack it is necessary to find the gradient of the loss function with respect to changing the $\mathbf{x}'$ parameters. When calculating the $\boldsymbol \theta$ gradients during the FL model training, the parameter shift rule can be used to find the gradients,
 \begin{equation}
    \begin{split}
            \frac{\partial \texttt{Cost}(\boldsymbol \theta)}{\partial \theta_j} &= -2 \left(y -y(\boldsymbol \theta)\right) \frac{\partial y(\boldsymbol \theta)}{\partial \theta_j} \\
    &= -\left(y -y(\boldsymbol \theta)\right) (y(\boldsymbol \theta^+) - y(\boldsymbol \theta^-))
    \end{split}
\end{equation}

To find the derivatives of these gradients with respect to $\mathbf{x}'$ parameters we can similarly apply the parameter shift rule,

\begin{equation}
    \begin{split}
            &\frac{\partial}{\partial x'_k}\frac{\partial \texttt{Cost}(\boldsymbol \theta)}{\partial \theta_j}
    = \frac{\partial y_i(\boldsymbol \theta)}{\partial x'_k} (y(\boldsymbol \theta^+) - y(\boldsymbol \theta^-)) \\
    &\hspace{26mm}+ y(\boldsymbol \theta) (\frac{\partial y(\boldsymbol \theta^+)}{\partial x'_k} - \frac{\partial y(\boldsymbol \theta^-)}{\partial x'_k})\\
    &= \frac{1}{2}\Big( \big( y(\boldsymbol x'^+, \boldsymbol \theta) - y(\boldsymbol x'^-, \boldsymbol \theta) \big) \big(y(\boldsymbol x', \boldsymbol \theta^+) - y(\boldsymbol x', \boldsymbol \theta^-) \big) \\
    & + y(\boldsymbol x', \boldsymbol \theta) \big( y(\boldsymbol x'^+, \boldsymbol \theta^+) - y(\boldsymbol x'^-, \boldsymbol \theta^+) - y(\boldsymbol x'^+, \boldsymbol \theta^-)  \\
    & + y(\boldsymbol x'^-, \boldsymbol \theta^-) \big) \Big)   
    \label{Eq:cost_ms_grad_diffeentiated}
    \end{split}
\end{equation}
noting that additional circuit evaluations will be required to compute this quantity as this is a second-order partial derivative. In the partial derivative with respect to $\theta_j$ and $x'_k$ the parameter shift variables $\boldsymbol \theta ^{\pm}$ and $\boldsymbol x'^{\pm}$ correspond to taking $\theta_j \rightarrow \theta_j \pm \frac{\pi}{2}$ and $x'_k \rightarrow x_k \pm \frac{\pi}{2}$ respectively. It is then possible to calculate the gradient of the loss function,
\begin{equation}
\begin{split}
     \frac{\partial L}{\partial x'_k} = \sum_{j=1}^d -2 \Big( \frac{\partial}{\partial x'_k} \frac{\partial \texttt{Cost}(\boldsymbol \theta, \mathbf{x'}, y)}{\partial \theta_j} \Big) \Big( \frac{\partial}{\partial \theta_j}\texttt{Cost}(\boldsymbol \theta, \mathbf{x}', y) \\
     - \frac{\partial}{\partial \theta_j}\texttt{Cost}(\boldsymbol \theta, \mathbf{x}, y) \Big)   
\end{split}
\end{equation}
by substituting in the parameter shift values for the first and second order partial derivatives of $\texttt{Cost}(\boldsymbol \theta, \mathbf{x}',y)$ along with the known client gradients. We can now utilize this to perform a gradient descent to match the known client gradients while optimizing $\mathbf{x'}$ in order to attempt to recover the target client's data $\mathbf{x}$. 

The above machine-learning attack involves the server using the same VQC architecture to compute the gradients with respect to $\mathbf{x}'$ in order to feed it to the above loss function. One can alternatively argue whether it is possible to build a purely classical machine learning attack based on the gradients received from the clients. Such a model would require learning the mapping from the gradient value $\frac{\partial}{\partial \theta_j}\texttt{Cost}(\boldsymbol \theta, \mathbf{x}', y)$ to the input value $\mathbf{x}'$. This would require training data comprising of gradient values alongside their true input values in order to learn the mapping. Generating this training dataset would then require evaluating gradients of quantum circuits many times in order to generate sufficient training data. 
Furthermore, we expect that the number of training samples required to learn this relationship would likely scale with the highest frequency term so as to not violate the Nyquist-Shannon Theorem \cite{shannon1949}. Therefore, there is no practical purely classical machine learning attack, as any classical algorithm will require sampling a quantum circuit potentially exponentially many times in order to initially train the classical model. 

\subsubsection{Privacy from machine learning attack}

Our key argument for hardness rests on the fact that as $m$ increases, the highest frequency per input dimension in the Fourier series increases exponentially $d_F = \frac{5^m - 1}{2}$. This results in a loss function where the number of local minima points scale exponentially with $m$ such that the minima points are uniformly distributed on average across the loss landscape (See Fig.~\ref{fig:log_minima_vs_log_freqsize_univariate} and  Fig.~\ref{fig:dist_local_minima_from_global_minima} respectively). Hence a sufficiently high $m$ will generate a loss function where optimizations get stuck close to their initialization point, thus requiring an average number of re-initializations that scales exponentially in $m$ to find the global minimum. 

We also observe that the average distance between the global minimum and the nearest local maxima $r$ decreases exponentially with $m$ (see Fig.~\ref{fig:avg_spacing}). This means that even an ideal stochastic optimizer with the ability to break out of any local minima would need to limit the distance it jumps each time by an amount proportional to $r$, so as to avoid missing the global minimum, and hence the total number of stochastic jumps would scale exponentially with $m$. Both stochastic and non-stochastic gradient descent methods imply some repeated sampling (via average re-initializations, or average stochastic jumps, required to reach the global minimum) that scales exponentially as the number of qubits per input $m$ is increased. 

The high-frequency terms make the gradient inversion attack loss landscape hard to train as we vary $\mathbf{x}'$. However, this does not affect the original FL model training since the original model is overparameterized in $\boldsymbol \theta$, in contrast to the attack model which is severely \emph{underparameterized} in $\mathbf{x}'$. Due to the underparamterization of the highly expressive attack model, the local minima are spaced uniformly throughout the loss landscape, making it extremely hard to find the global minimum. Equally spaced local minima also implies that even if the loss function of the gradients is minimized such that $L(\boldsymbol \theta, \mathbf{x'}, y) < \epsilon_L$ (assuming a local minima has loss function value below $\epsilon_L$), then while this means it may have found a good fit for the gradients, it could still be the case that $\rvert  \mathbf{x'} -  \mathbf{x} \rvert$ has not converged below a threshold level $\epsilon$ and the attack model has failed to recover the client input. This decorrelation between optimizing gradients and recovering user data is shown in Fig.~\ref{fig:dist_local_minima_from_global_minima}.

\subsubsection{Theoretical bound of local minima}

With the distinction made between an underparameterized attack model and an overparameterized FL global model, we can focus on exploring bounds on the number of local minima in the attack model, as a measure of how difficult the machine-learning based gradient inversion attack will be. Local minima will occur in the loss function when the gradient of the loss with respect to $\mathbf{x}'$ is equal to zero i.e.,
\begin{equation}
\begin{split}
    \frac{\partial L}{\partial x'_k} = & \sum_{j=1}^d -2 \Big( \frac{\partial}{\partial x'_k} \frac{\partial \texttt{Cost}(\boldsymbol \theta, \mathbf{x}', y)}{\partial \theta_j} \Big) \Big( \frac{\partial}{\partial \theta_j}\texttt{Cost}(\boldsymbol \theta, \mathbf{x}', y) \\
    &- \frac{\partial}{\partial \theta_j}\texttt{Cost}(\boldsymbol \theta, \mathbf{x}, y) \Big) = 0
\end{split}    
\end{equation}

We previously identified that the $\texttt{Cost}$ could be expressed in Chebyshev polynomials with $2n$ unknowns and with max degree $(d_F)^n=(5^{m}-1)^n/2^n$. Therefore the corresponding polynomial for $\frac{\partial L}{\partial x'_k}$ will have maximal degree $2(d_F)^n$. In the case of finitely many local minima, one can utilize B\'{e}zout's theorem which states that the number of roots of a system of polynomial equations in the non-degenerate case can be bounded by the product of the degree of the polynomials \cite{cox2013ideals}, 
\begin{equation}
    2^n \texttt{deg}\left[\frac{\partial L}{\partial x'_1}\right]...\texttt{deg}\left[\frac{\partial L}{\partial x'_n}\right] = 4^n \left( \frac{5^{m}-1}{2} \right) ^{n^2}
\end{equation}
Hence the number of local minima is upper bounded by this quantity \cite{you2021exponentially}. It should be noted that this result is for a general system of polynomials. In practice, we find that the average amount of local minima in the loss landscape seems to arise from the highest frequency term in the Fourier series. This seems intuitive considering a model with only a single frequency $\omega$, for example the function $\cos (\omega x)$, would contain $\omega$ local minima in the range $x \in [0,2\pi]$. Section~\ref{sec:numerics} details our numerical results which suggest that on average the gradient inversion attack loss landscape follows a similar pattern.

\subsubsection{Solutions to systems of equations in feature space} \label{sec:feature_space}

In this section, we explore the feasibility of solving the system of equations generated when the gradients of the VQC model is written as a system of equations in the feature space, i.e., as a function of the quantum state $\ket{\psi(\mathbf{x})} = U(\gamma \mathbf{x})\ket{0}$, instead of directly expressing it as a function of the input $\mathbf{x}$ as we have been analyzing in the previous sections. An immediate observation when directly dealing in the feature space is that, the quantum model output is linear in the $\rho(\mathbf{x}) = \ket{\psi(\mathbf{x})}\bra{\psi(\mathbf{x})}$ and is expressed as,
\begin{equation}
    y(\mathbf{x}, \theta) = \text{Tr}[\mathbf{O}W(\boldsymbol\theta) \rho(\mathbf{x}) W^\dagger (\boldsymbol\theta)]
\end{equation}

Given that $\rho(\mathbf{x})$ has no $\boldsymbol\theta$ dependency, we can write this out explicitly in matrix elements as
\begin{equation}
    y(\mathbf{x}, \boldsymbol\theta) = \sum_{i,l,k,m}\mathbf{O}_{il}W_{lk} W^\dagger_{mi} \rho(\mathbf{x})_{km}
\end{equation}
We can see from this that, similar to the dense linear classical neural network case where the model output has a linear form with respect to the input, the quantum model output has a linear form in the feature space of $\rho(\mathbf{x})$, in contrast to the input $\mathbf{x}$ space. The degree of the polynomials of the gradient equations in the feature space will be dependent on the exact cost function used. For example the cost function of the form mean squared error loss will give rise to the gradients of the form,
\begin{equation}
\begin{split}
    \frac{\partial}{\partial \theta_j}\texttt{Cost} &= 2(y -y(\mathbf{x}, \theta)) \frac{\partial}{\partial \theta_j}y(\mathbf{x}, \theta)   \\
    & = 2(y-\sum_{i,l,k,m}\mathbf{O}_{il}W_{lk} \rho(\mathbf{x})_{km} (W^\dagger)_{mi})\\
    &*(\sum_{i',l',k',m'}\mathbf{O}_{i'l'} \rho(\mathbf{x})_{k'm'} \frac{\partial}{\partial \theta_j}(W_{l'k'}(W^\dagger)_{m'i'})))
\end{split}
\end{equation}
Thus, the resulting gradient expressions are quadratic in the elements of $\rho(\mathbf{x})$. Hence with sufficiently many gradients (which will be the case if using an overparameterized FL model), one would be able to reconstruct $\rho(\mathbf{x})$ by solving a system of quadratic equations, which would be easier than solving for $\mathbf{x}$ directly as examined in previous sections. This is a more similar comparison to the classical neural network case.

However, finding the form of the equations to be solved in feature space will in general require explicitly finding the exact form of $W(\boldsymbol\theta)$ in order to perform the appropriate matrix multiplications. This task will scale exponentially in general, in the number of qubits, as was also the case for finding an analytical form of the coefficients in the previous sections.

While solving the system of equations to find $\rho(\mathbf{x})$ may be easier than for $\mathbf{x}$, we would then also be faced with the task of retrieving $\mathbf{x}$ from $\rho(\mathbf{x})$. The encoding circuit we propose does give a bijective feature map $\mathbf{x} \rightarrow \ket{\psi(\mathbf{x})}$, as long as $\gamma\mathbf{x}$ is between $[0, 2\pi)$. However, even given a quantum state $\ket{\psi(\mathbf{x})}$ and knowing the encoding circuit that generates it, it is not possible to find the inverting circuit unless one knows $\mathbf{x}$ to begin with. To acquire $\mathbf{x}$ one method would be to calculate the overlap $\braket{\psi(\mathbf{x})}{\psi(\mathbf{x}')} = 1$ by varying parameters $\mathbf{x}'$. The problem has been effectively transformed into passing $\mathbf{x}'$ through some feature map, and trying to match it up with the known output of the feature map when $\mathbf{x}$ is passed through it.

If we wish to maximize the overlap $\braket{\psi(\mathbf{x})}{\psi(\mathbf{x}')}$ though appropriate cost function minimization, we would again encounter the problem of exponentially many well separated local minima in the loss landscape due to the high expressivity of the encoding map as discussed previously. Due to the simplicity of product feature map we have used to encode the classical data, it is possible in this specific case to write this out in an analytical form (this is not the case in general for a more complicated quantum feature maps). Consider the Fourier tower map for a univariate input $x$ described previously using $R_X(\gamma x)$ gates which produces the state $\ket{\psi(x)} = U(\gamma x)\ket{0}$,
\begin{equation}
\begin{split}
    &(\cos(\gamma x)\ket{0} + \sin(\gamma x)\ket{1})\otimes(\cos(5\gamma x)\ket{0} + \sin(5\gamma x)\ket{1})  \\
    & \otimes \cdots \otimes \cos(5^{m-1}\gamma x)\ket{0} + \sin(5^{m-1}\gamma x)\ket{1})
\end{split}
\end{equation}
Clearly each $x$ will give a unique $\ket{\psi(x)}$ as can be seen by looking at the first qubit only. To calculate the overlap between two states we want to find $\braket{\psi(x)}{\psi(x')} = \bra{0}U^\dagger(\gamma x')U(\gamma x)\ket{0} = 1$. We need to find $x'$ for which the probability of measuring $\ket{0}$ on all qubits in the circuit $U^\dagger(\gamma x')U(\gamma x)\ket{0}$ is equal to one. If $U(\gamma x)$ is known to be the previously introduced Fourier tower map encoding using $R_X$ gates then we can write the probability of measuring zeros on all qubits as,
\begin{equation}
    \rvert \cos(\gamma x-\gamma x') \cos(5(\gamma x- \gamma x')) \cdots \cos(5^{m-1}(\gamma x- \gamma x')) \rvert^2
\end{equation}
and we need to find the maximal value of this probability, which will be the case when the function equals $1$. This equation in general is a high degree polynomial that must be solved, where the highest degree will be exponential $m$ when writing out in Chebyshev form, and hence by previous arguments the difficulty will scale at least exponentially in $m$. Likewise, we can easily see that any attempt at performing any form of gradient descent to find a numerical solution will be swamped with exponentially many local minima due to its periodic nature. These local minima will be uniformly distributed through this cost function and the spacing between these minima will be inversely proportional to $5^{m-1}$. It is exactly these properties that manifest also in Sec.~\ref{sec:numerics}, although a simple analytical form in the input space is not possible to show explicitly. The system of equations in the feature space are exponentially hard to write down, requiring unitary simulation of the overparameterized variational circuit. If they are written down, they may easier to solve in order to find $\rho(x)$, when compared to trying to find $x$ directly. However, even if $\rho(x)$ is found from these equations, then finding the $x'$ that inverts the state $\ket{x}$ will similarly be exponentially hard to solve. Because of several exponentially scaling steps, we will not consider this as a viable attack method and will move on to direct gradient inversion machine learning based attacks, which do not require formulating a system of equations and hence skip this exponentially hard step. However, considering the problem of trying to learn $x$ given $\ket{x}$ in this example explicitly shows how local minima would appear, and how they would cause machine learning optimization algorithms exponentially long to solve in terms of $m$. This property will reappear numerically in the subsequent machine learning gradient inversion attack.

\section{Numerical Results} \label{sec:numerics}

This section highlights the numerical results solidifying our privacy arguments. We highlight the results involving the machine learning gradient attacks, followed by the analysis on the loss landscape of the attack model. 

Taking the scaling factor $\gamma = 2\pi$ into account, $x$ and $x'$ values are normalized in the range $[0, 2\pi]$ for univariate input (and similarly for higher dimensions) in order to cover all unique quantum states when input as rotation angles. Therefore, in these numerical results we report the distance between two values using the closest angular distance between them on a circle defined as,
\begin{equation}
    \rvert x - x' \rvert  = \min{( \texttt{abs}(x - x'), 2\pi - \texttt{abs}(x - x') )}
\end{equation}

\subsection{Machine learning gradient attack result}

\begin{figure}[t]
\includegraphics[scale=0.44]{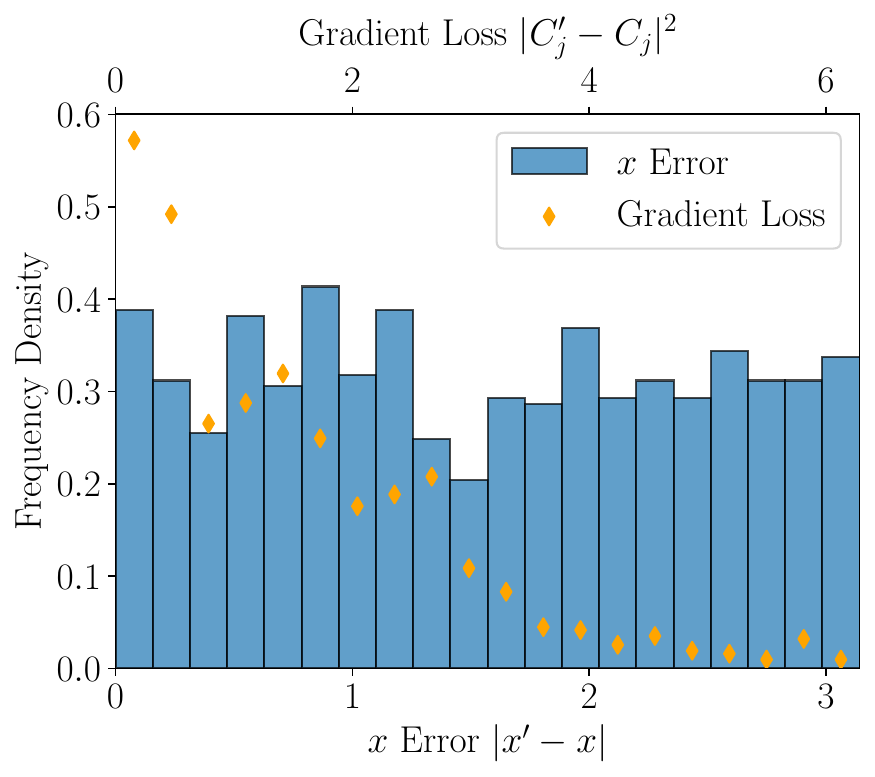}
\caption{(Blue) Histogram of the error between client input $x$ and attack prediction value $x'$ for VQC model with $m=3$ qubits after 60 iterations of ADAM optimizer. (Orange) The loss value between predicted gradient and true gradient $|C_j' - C_j|^2$. Results are for fixed randomly chosen $\boldsymbol\theta$, while $x$ was randomly initialized over $100$ experiments and within each experiment, $10$ attempts were made with different $x'$ initializations.}
\label{fig:66_weights_adam_opt_distribution}
\end{figure}

\begin{figure}[t]
\centering
\includegraphics[scale=0.035]{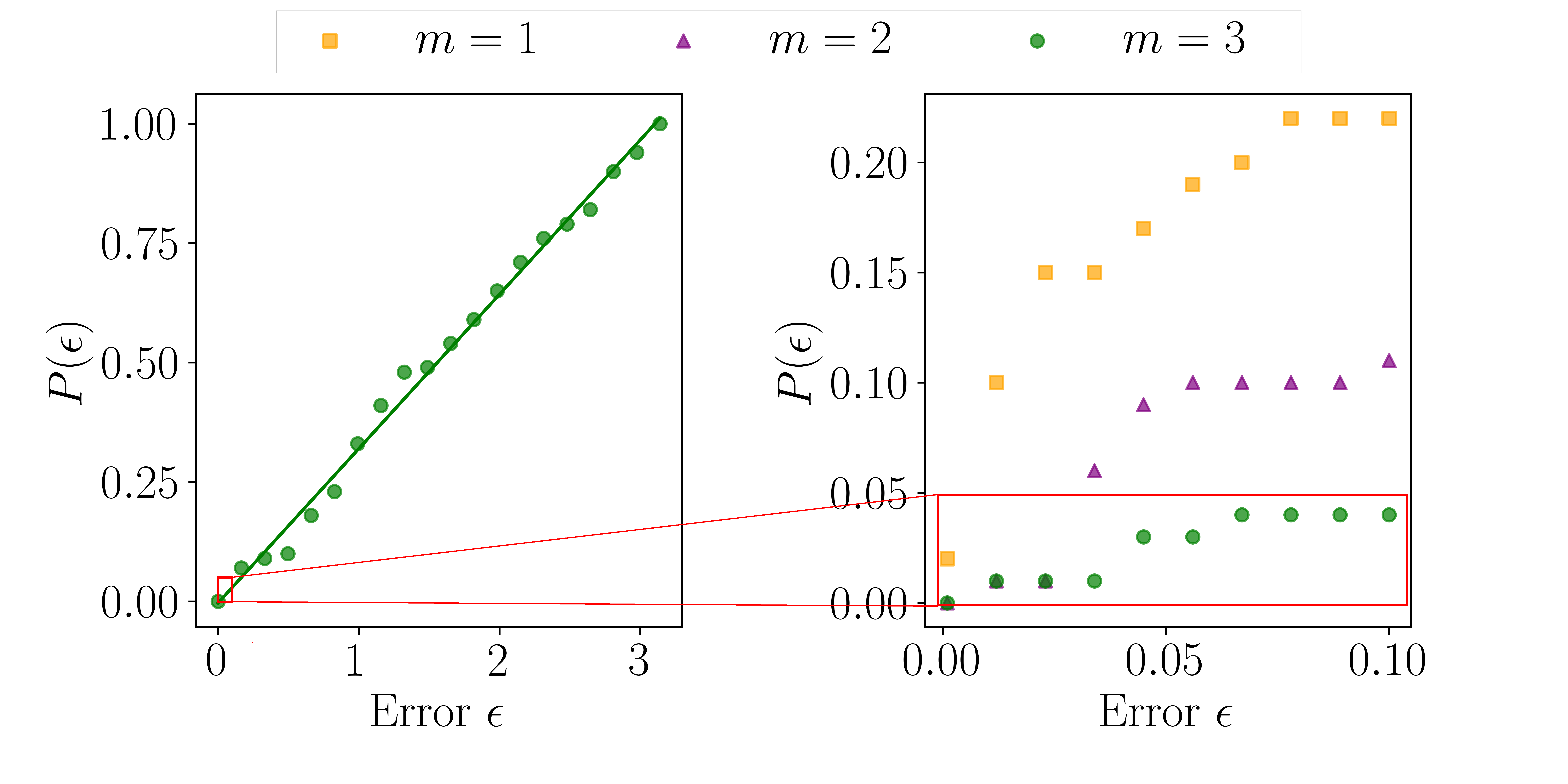}
\caption{The probability of a successful attack $P(\epsilon)$ for a given initialization as the acceptable error $\epsilon$ is varied. Attack performed using gradient descent and ADAM optimization with learning rate $\eta=0.01$. (Left) Large scale plot of the model when $m=3$ demonstrating the probability of success is linearly correlated with $\epsilon$. (Right) Results for different models with varying $m$ for $0 \leq \epsilon \leq 0.1$, demonstrating the lower $m$ values exhibit worse privacy.}
\label{fig:probability_versus_error_small_scales}
\end{figure}

Utilizing gradient descent via the ADAM optimization method for the learning rate $\eta = 0.01$,  we attempted the gradient attack on a univariate model i.e., $n = 1$ with $m=3$. The resulting distribution of the distance between the final $x'$ and client data $x$ is shown in Fig.~\ref{fig:66_weights_adam_opt_distribution}. The distribution generated does not show a peak around $\rvert x - x' \rvert = 0$ and in fact appears uniformly distributed. This implies there is a sufficient amount of local minima distributed such that the probability of convergence effectively depends on the random initialization. Further evidence for this conclusion is provided in Figure~\ref{fig:probability_versus_error_small_scales} which shows there is a linear relationship between error tolerance $\epsilon$ and the probability of a successful attack, where success is defined as achieving $\rvert x-x'\rvert < \epsilon$, with a coefficient of determination $R^2 = 0.993$. This linear relationship is what would be expected from a model guessing $x'$ randomly. Hence running ADAM optimization for $m = 3$ is no better than guessing at random. Note there may be other stochastic optimization methods that have a higher probability of reaching a solution eventually, however, they would still need to be limited by the width of the valley of the global minimum, which is shown to shrink exponentially as $m$ increases in Fig.~\ref{fig:avg_spacing}.

\subsection{Analysis of loss landscape}

In order to analyze the loss landscape of the gradient attack, we used simulations in Qiskit \cite{Qiskit} to calculate the entire loss landscape, for an appropriate sampling granularity, from which the location of all critical points could be identified. This was done by randomly fixing the $\boldsymbol\theta$ parameters and client data $\mathbf{x}$ in the range $[0, 1]$, leading to $\gamma\mathbf{x} \in [0, 2\pi]$, and the client output $y \in \{-1, 1\}$. We then calculate the loss of the difference of estimated and target gradients using $l_2$ error. Instead of taking the sum of the $l_2$ error of all gradients, in practice, it suffices to find the global minimum of a single gradient,
\begin{equation}
    L_j = \left(\frac{\partial}{\partial \theta_j}\texttt{Cost}(\boldsymbol \theta, \mathbf{x}', y) - \frac{\partial}{\partial \theta_j}\texttt{Cost}(\boldsymbol \theta, \mathbf{x}, y)\right)^2
\end{equation}
as the global minimum obtained from matching a single gradient can easily be verified against all other gradients to check if $L_j = 0, \forall j \in [d]$. This significantly reduces the number of circuit evaluations required, especially as we consider circuits that are overparameterized in $\boldsymbol\theta$ in this investigation. In these numerical results, the gradient index $j$ was chosen randomly from all possible $\boldsymbol\theta$ values, excluding the values in the very final layer of the circuit.

\begin{figure}[t]
\centering
\includegraphics[scale=0.44]{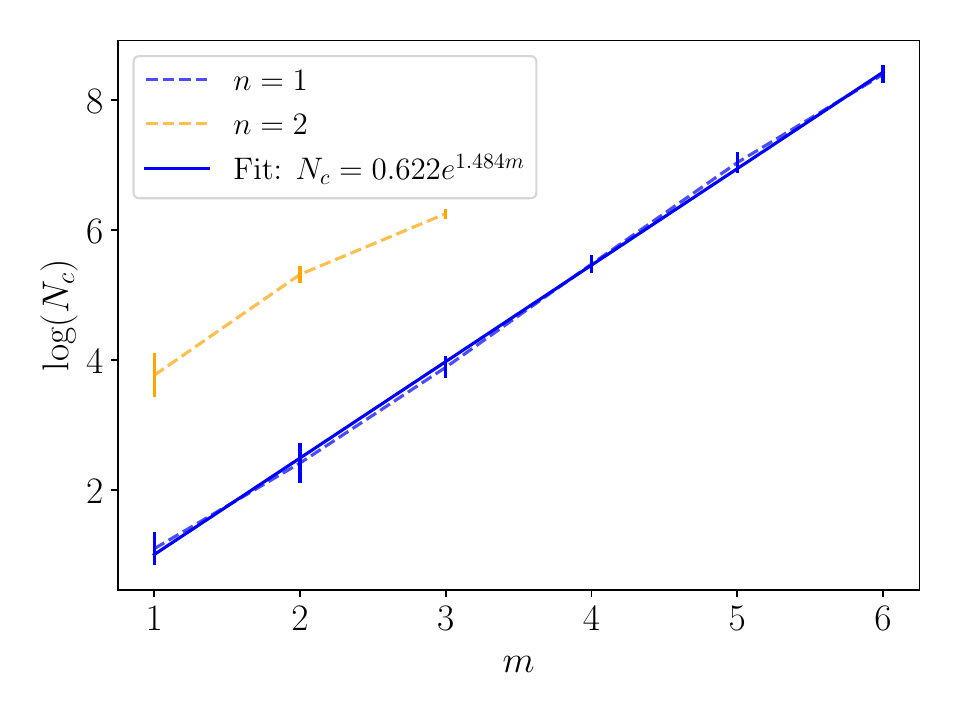}
\caption{The average logarithm of number of local minima $N_c$ plotted against the number of qubits per input $m$. The error bars correspond to the standard deviation over $10$ repeated experiments with different randomized parameters.}
\label{fig:log_minima_vs_log_freqsize_univariate}
\end{figure}

\begin{figure}[h!]
\includegraphics[scale=0.29]{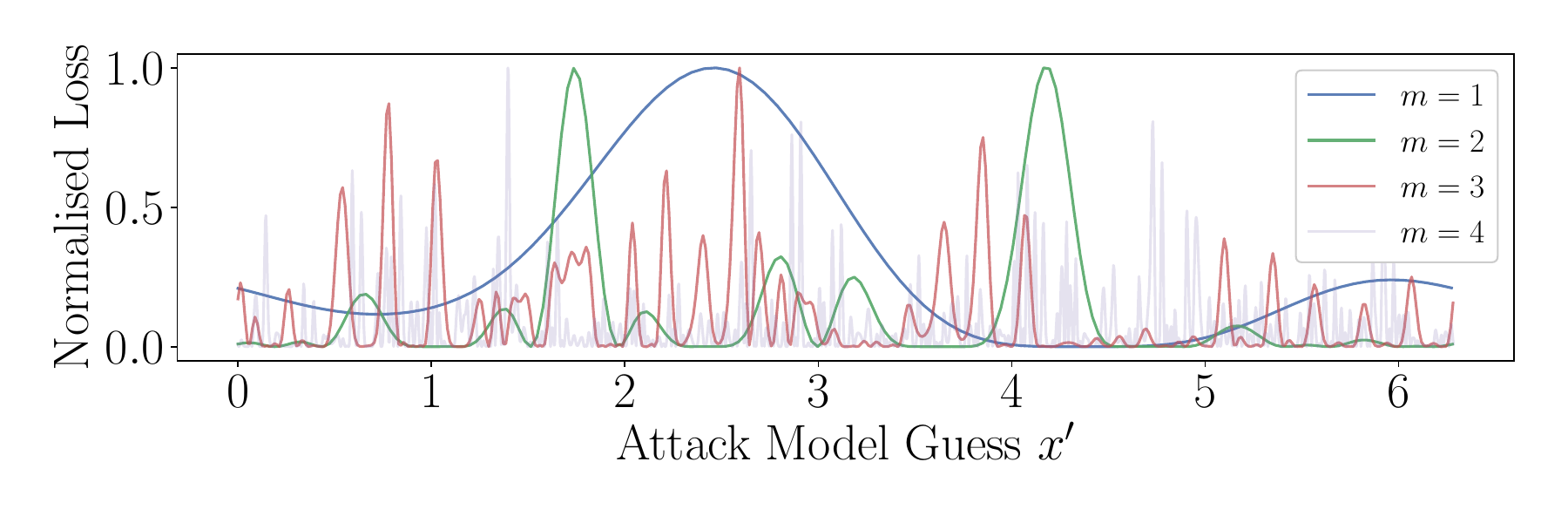}
\caption{Visualization of the loss landscape of the gradient inversion model attack as the attack model guess $x'$ is varied in the univariate case $n=1$. The loss is plotted for models with differing numbers of qubits per input $m$. The shape of the loss landscape is influenced by the highest frequency terms which scale exponentially with $m$. The plotted loss is normalized for readability.}
\label{fig:univariate_loss_landscape_comparison}
\end{figure}

\begin{figure}[h!]
    \centering
    \begin{subfigure}{0.25\textwidth}
        \includegraphics[width=\textwidth]{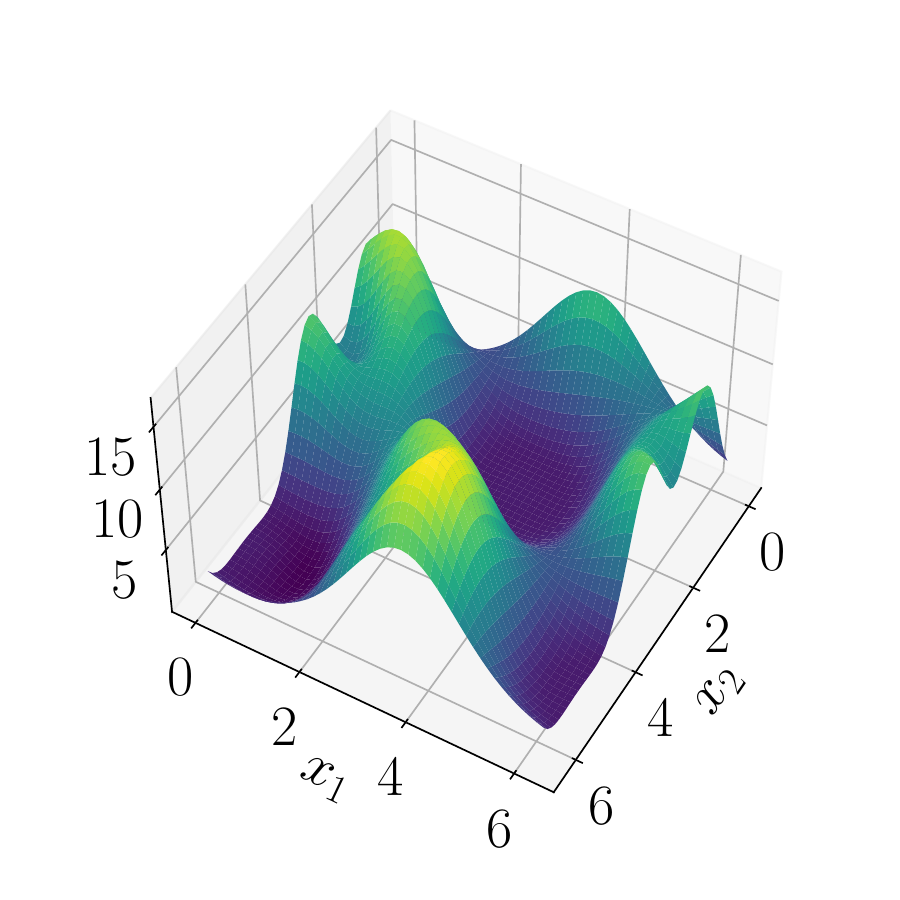}
    \end{subfigure}
    \begin{subfigure}{0.21\textwidth}
        \includegraphics[width=\textwidth]{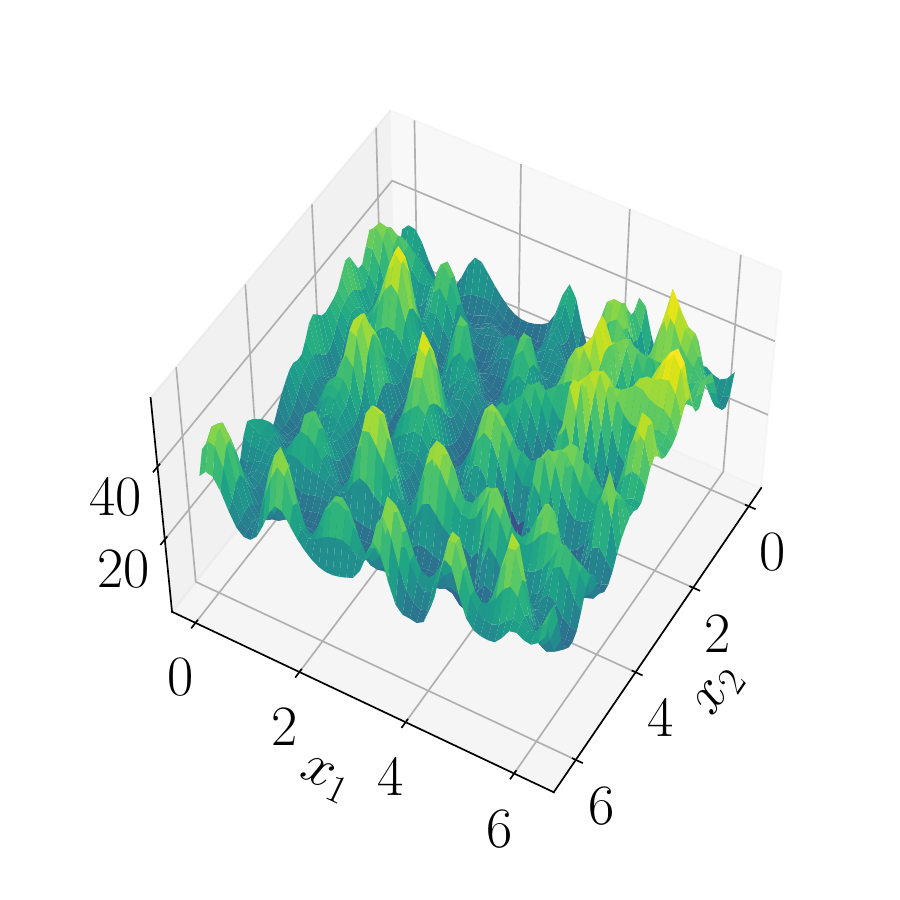}
    \end{subfigure}
    \caption{Gradient inversion attack loss landscape for (Left) $m=1$ and (Right) $m=2$ for the multivariate case $n=2$ with two inputs $\mathbf{x}' = (x_1, x_2)$.}
    \label{fig:multivariate_loss_landscape_comparison}
\end{figure}

Fig.~\ref{fig:log_minima_vs_log_freqsize_univariate} demonstrates that the number of local minima in the loss function for the univariate case scales as,
\begin{equation}
    N_c = 0.622e^{1.484m}
\end{equation}
where $N_c$ is the number of local minima points. Even the two dimensional case exhibits a pattern of local minima scaling as exponential with the number of qubits per encoding $m$. Example visualizations of how the loss landscape changes as $m$ is varied can be seen in Fig.~\ref{fig:univariate_loss_landscape_comparison} for the univariate case $n=1$ and in Fig.~\ref{fig:multivariate_loss_landscape_comparison} for the multivariate case $n=2$.

\begin{figure}[h!]
\centering
\includegraphics[scale=0.46]{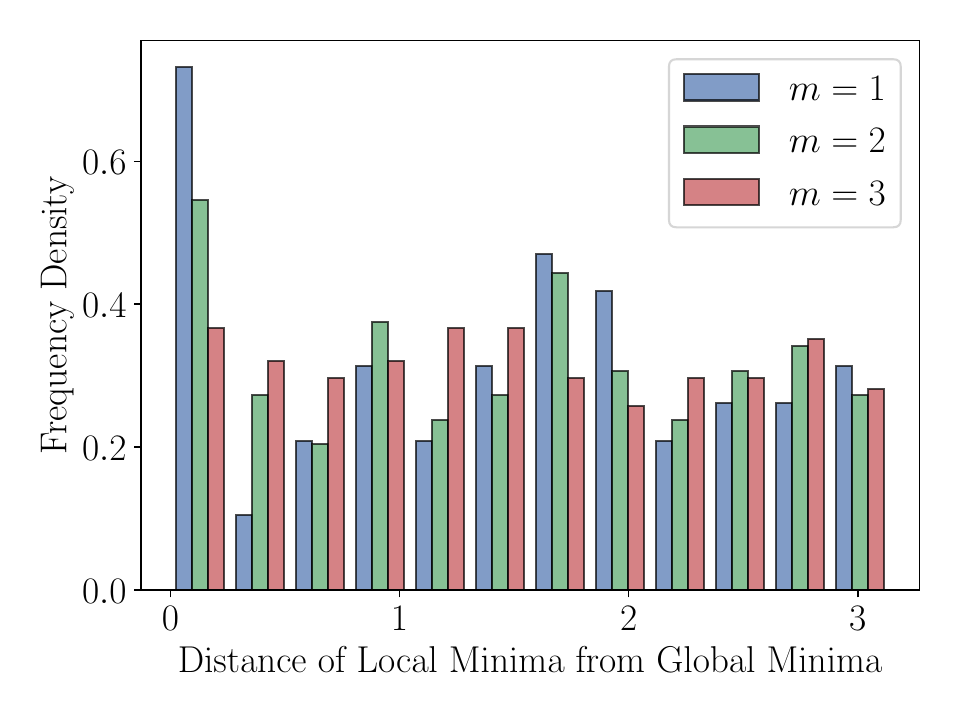}
\caption{Distribution of the distance of minima from the global minimum in the gradient attack loss landscape for differing amounts of qubits per input $m$. Results were sampled over $10$ repeated experiments with different randomized parameters.}
\label{fig:dist_local_minima_from_global_minima}
\end{figure}

A large number of local minima does not immediately guarantee that the model is difficult to train. Overparamaterized quantum models have been shown to contain local minima distributed exponentially close to the global minimum \cite{anschuetz2022} which would not pose much challenge to an attack. In the setting of gradient attack however, we are firmly in the underparameterized regime as we have $n$ inputs $\mathbf{x'} = x'_1\cdots x'_n$ to vary but $N_q = nm$ qubits in total. The results shown in Fig.~\ref{fig:dist_local_minima_from_global_minima} demonstrate the distribution of local minima throughout the gradient attack loss function. While for $m=1$ and $m=2$ the overall distribution is mostly uniform, we can see there is a slightly elevated density of minima close to the global minimum. However, by the time we reach $m=3$ and onwards, the distribution appears uniform.

\begin{figure}[h!]
\centering
\includegraphics[scale=0.44]{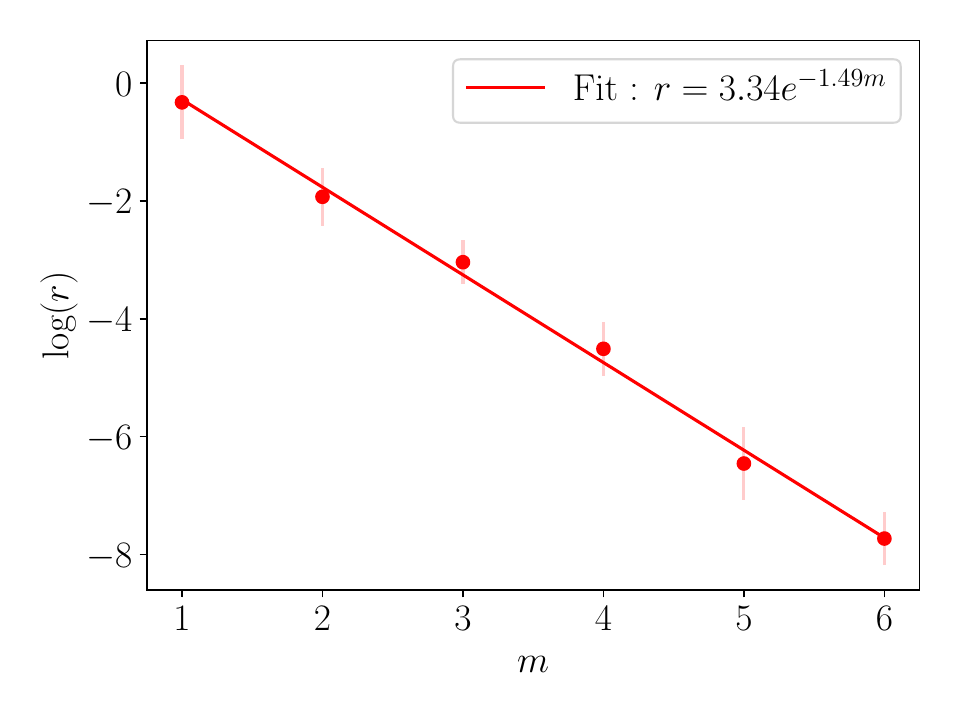}
\caption{The average logarithm of the spacing $r$ between the global minimum and the nearest local maxima plotted against the number of qubits per input $m$ for the univariate case $n=1$. The error bars correspond to the standard deviation over $10$ repeated experiments with different randomized parameters. This spacing defines a scale in which attack model initialization must be binned, or conversely the scale which limits the jump length of any stochastic optimization on the loss surface.}
\label{fig:avg_spacing}
\end{figure}

We see in Fig.~\ref{fig:avg_spacing} that the distance $r$ between the global minimum and the next local maxima decreases exponentially as $m$ increases, i.e., for $n=1$ the scaling is,
\begin{equation}
    r = 3.34e^{-1.49m}
\end{equation}
This is important as it gives an indication of the spacing of random initialization that would be required to guarantee convergence to the global minimum. Additionally, in the case of stochastic gradient descent it also gives an indication of the scale of how far the stochastic term can jump without missing the global minimum. This is the source of privacy against machine learning gradient inversion attacks, as on average both stochastic and non-stochastic gradient descent algorithms will have to sample a total number of points that scales exponentially with $m$. This provides an accessible method of increasing the privacy of the model.

\subsection{Attack model loss landscape analysis as original model is trained}

\begin{figure}[t]
    \centering
    \begin{subfigure}{0.48\textwidth}
        \includegraphics[width=\textwidth]{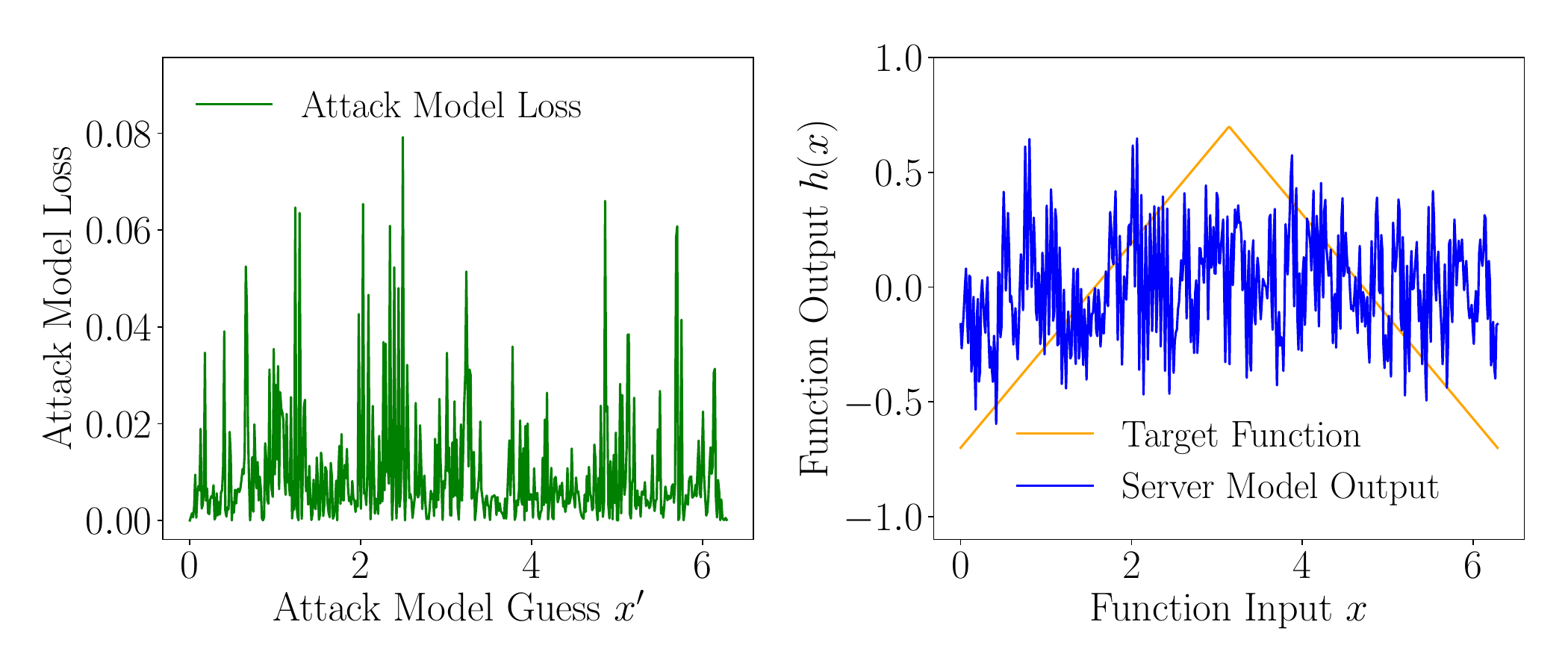}
        \caption{Attack loss and FL model fit at epoch $t = 0$}
    \end{subfigure}
    \begin{subfigure}{0.48\textwidth}
        \includegraphics[width=\textwidth]{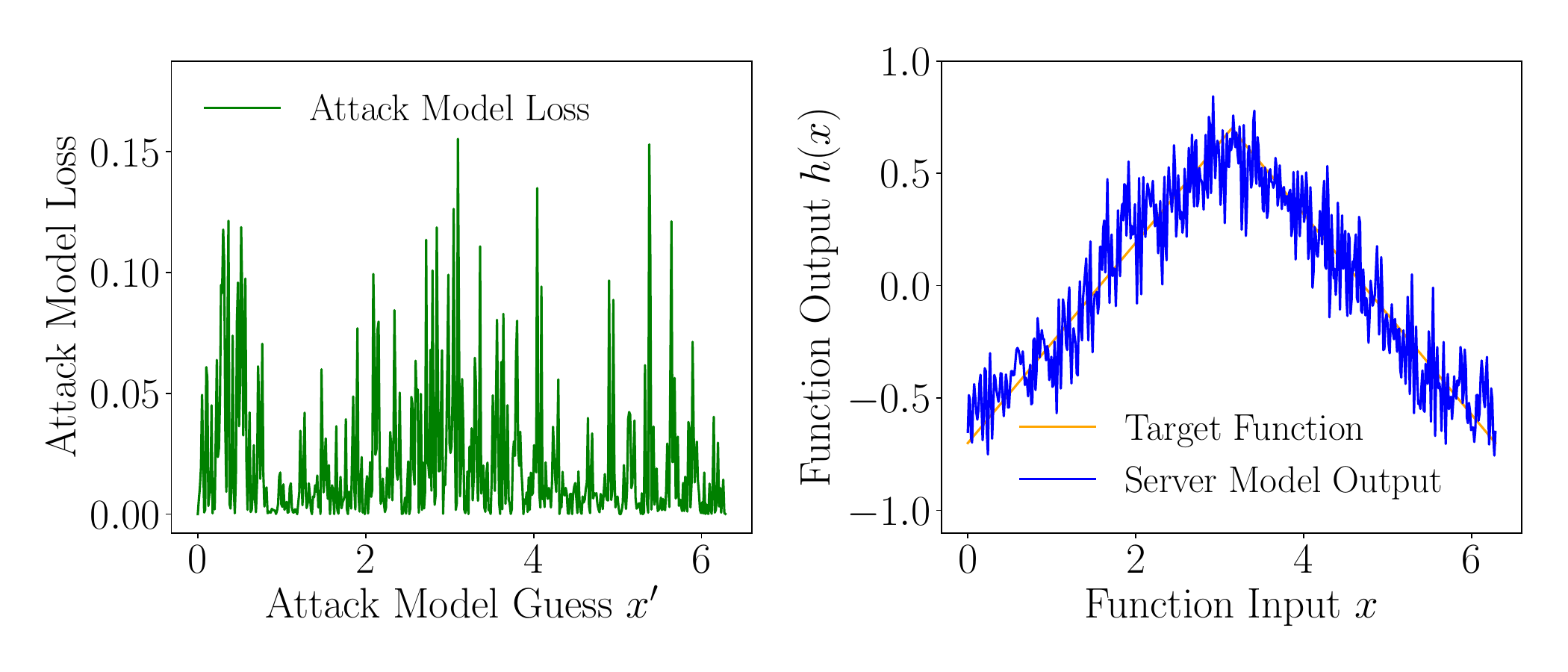}
        \caption{Attack loss and FL model fit at epoch $t = 20$}
    \end{subfigure}
    \begin{subfigure}{0.48\textwidth}
        \includegraphics[width=\textwidth]{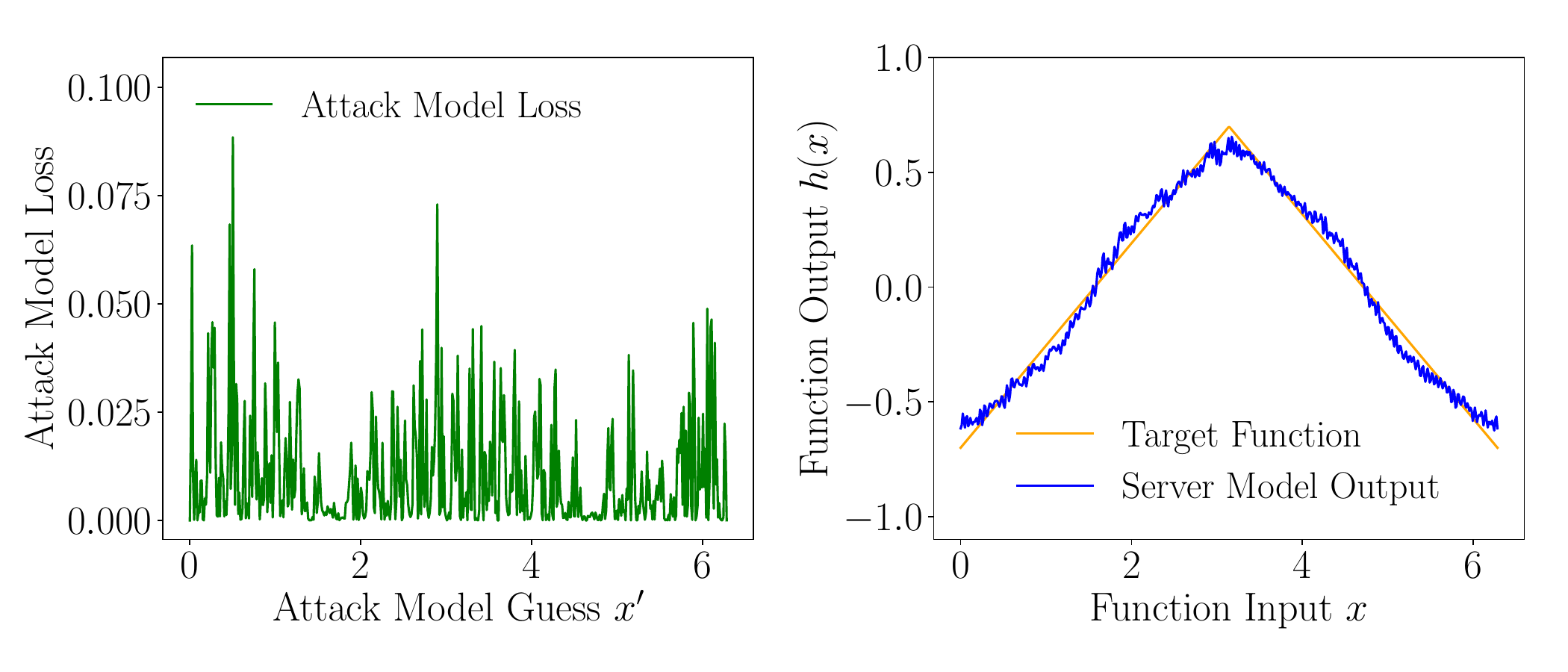}
        \caption{Attack loss and FL model fit at epoch $t = 400$}
    \end{subfigure}    
    \caption{(Left) The gradient attack model loss landscape at various points during the model's training. (Right) FL model output as it is trained to fit a target function. The loss landscape was calculated using the current thetas of the FL model, with the loss calculated using the gradients generated by $x=1.6371$ and $y=h(x)$. The model was univariate with a single $x$ input and four qubits per data point $m=4$. The FL model was fitted by taking training samples from the target function and training on them with a batch size 10 using ADAM optimizer with $\eta = 0.01$.}
    \label{fig:landscape_fitting_good}
\end{figure}

Next, we provide a visual analysis of how the loss landscape of the gradient attack model changes throughout the original training of the FL model, which corresponds to $\boldsymbol\theta$ values being updated each epoch as the FL model tries to fit some target function and hence the entire gradient attack loss function is changed at each step. In Fig.~\ref{fig:landscape_fitting_good} we show the gradient attack model loss function alongside the FL model output at various training epochs, as the FL model tries to fit a target function $h(x)$. Note that even though the target function in Fig.~\ref{fig:landscape_fitting_good} is relatively simple, straight lines require a large Fourier series with high degree frequencies to match well, hence the loss landscape maintains high-frequency terms and a gradient attack is hard to perform.

\subsection{Privacy when underlying FL model target function contains only low frequencies}

We have shown that the hardness of performing gradient attacks on this model derives from the large frequencies contained in the quantum model. For an arbitrary set of $\boldsymbol\theta$ the large frequency terms in the Fourier series that represents the quantum model can lead to a loss landscape that is difficult to train. The results reported so far assume $\boldsymbol\theta$ values are chosen randomly, or in the case of Fig.~\ref{fig:landscape_fitting_good} were trained to fit a function that has high degree frequency terms. This leads to the question of whether the privacy of the model would hold if the FL model is learning an underlying distribution that contains only low-frequency components. While in practice real-world data is not likely to exactly match some low-frequency signal, we will consider this case to bring extra insight into the model.

\begin{figure}[t]
    \centering
    \begin{subfigure}{0.48\textwidth}
        \includegraphics[width=\textwidth]{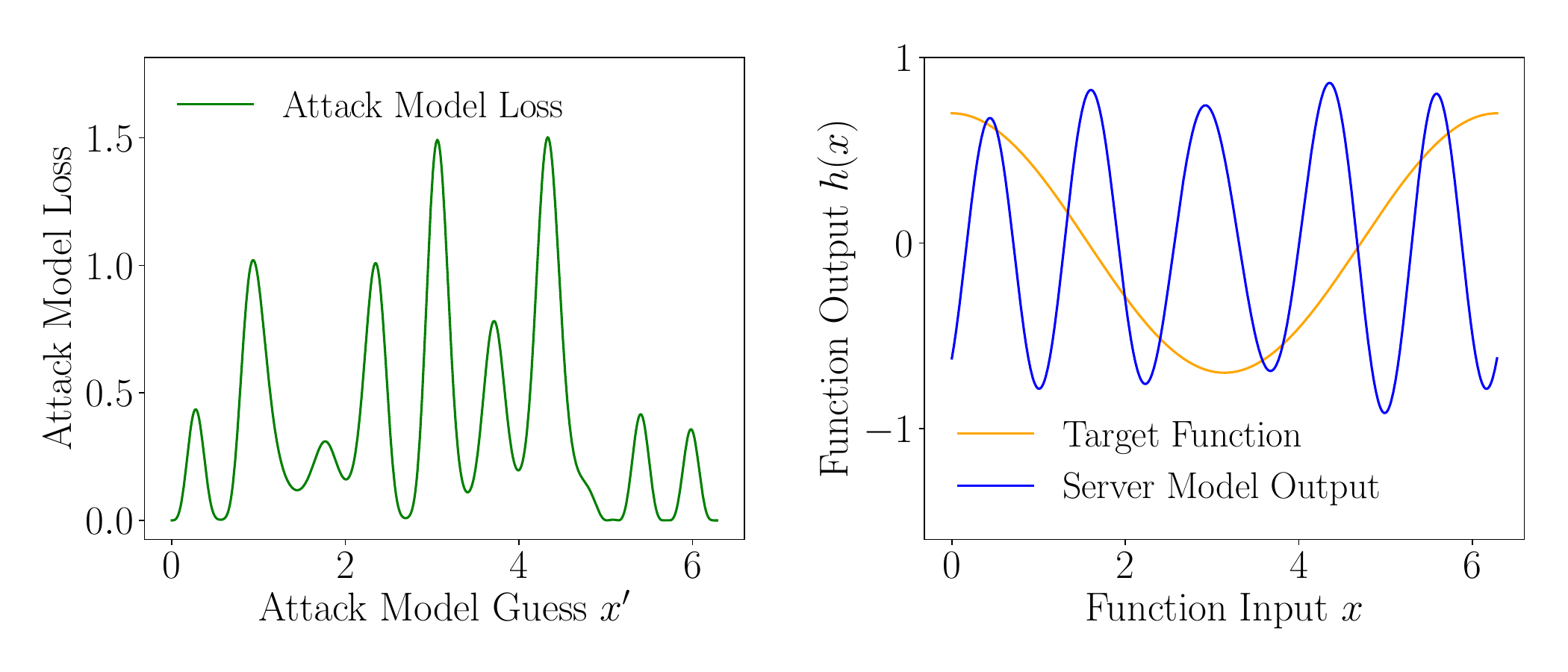}
        \caption{Attack loss and FL model fit at epoch $t = 0$}
    \end{subfigure}
    \begin{subfigure}{0.48\textwidth}
        \includegraphics[width=\textwidth]{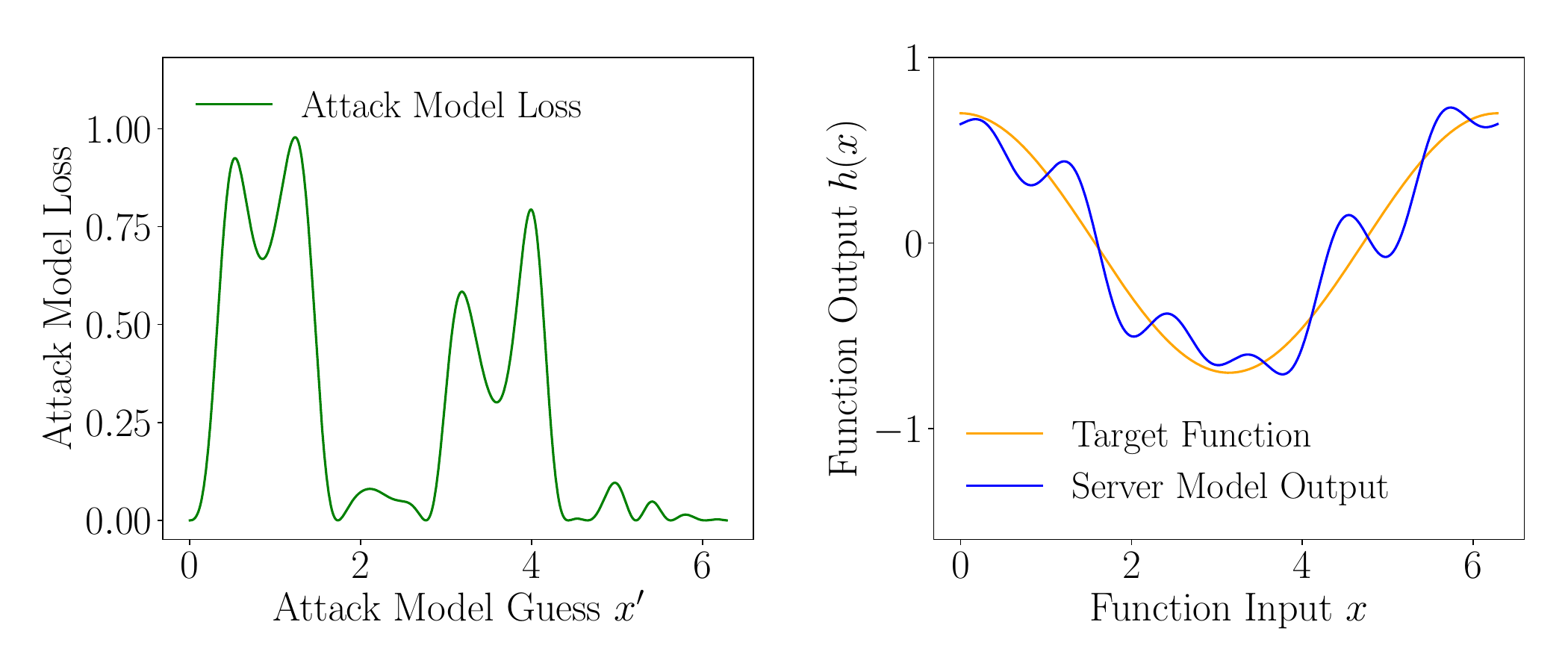}
        \caption{Attack loss and FL model fit at epoch $t = 100$}
    \end{subfigure}
    \begin{subfigure}{0.48\textwidth}
        \includegraphics[width=\textwidth]{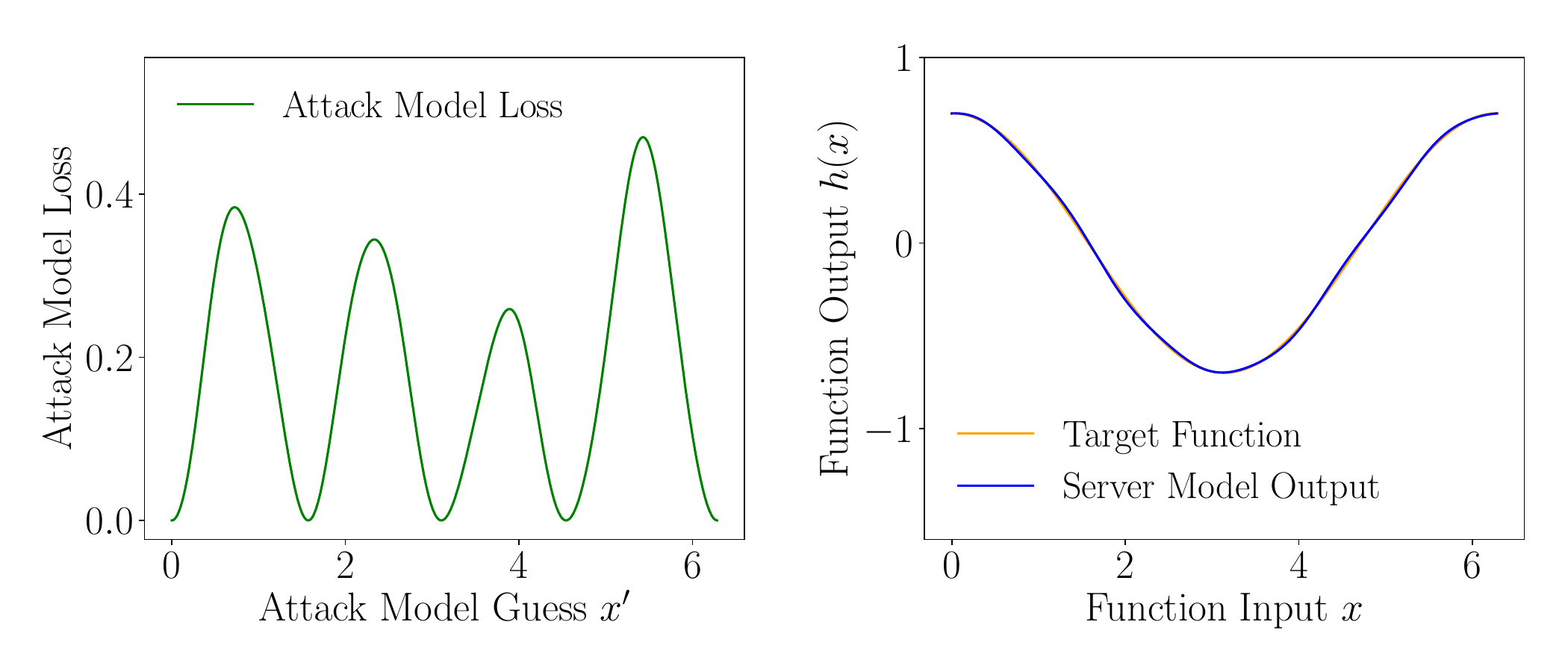}
        \caption{Attack loss and FL model fit at epoch $t = 400$}
    \end{subfigure}   
    \caption{(Left) The attack loss landscape when matching gradients during a gradient based attack at various points during the FL model's training. (Right) FL model output as it is trained to fit a simple cosine function $h(x)=0.7\cos {x}$. The loss landscape was calculated using the current thetas of the FL model, with the loss calculated using the gradients generated by $x=1.6371$ and $y=h(x)$. The model was univariate with a single $x$ input and two qubits per data point $m=2$.}
    \label{fig:landscape_flattening}
\end{figure}

\begin{figure}[h!]
    \centering
    \begin{subfigure}{0.48\textwidth}
        \includegraphics[width=\textwidth]{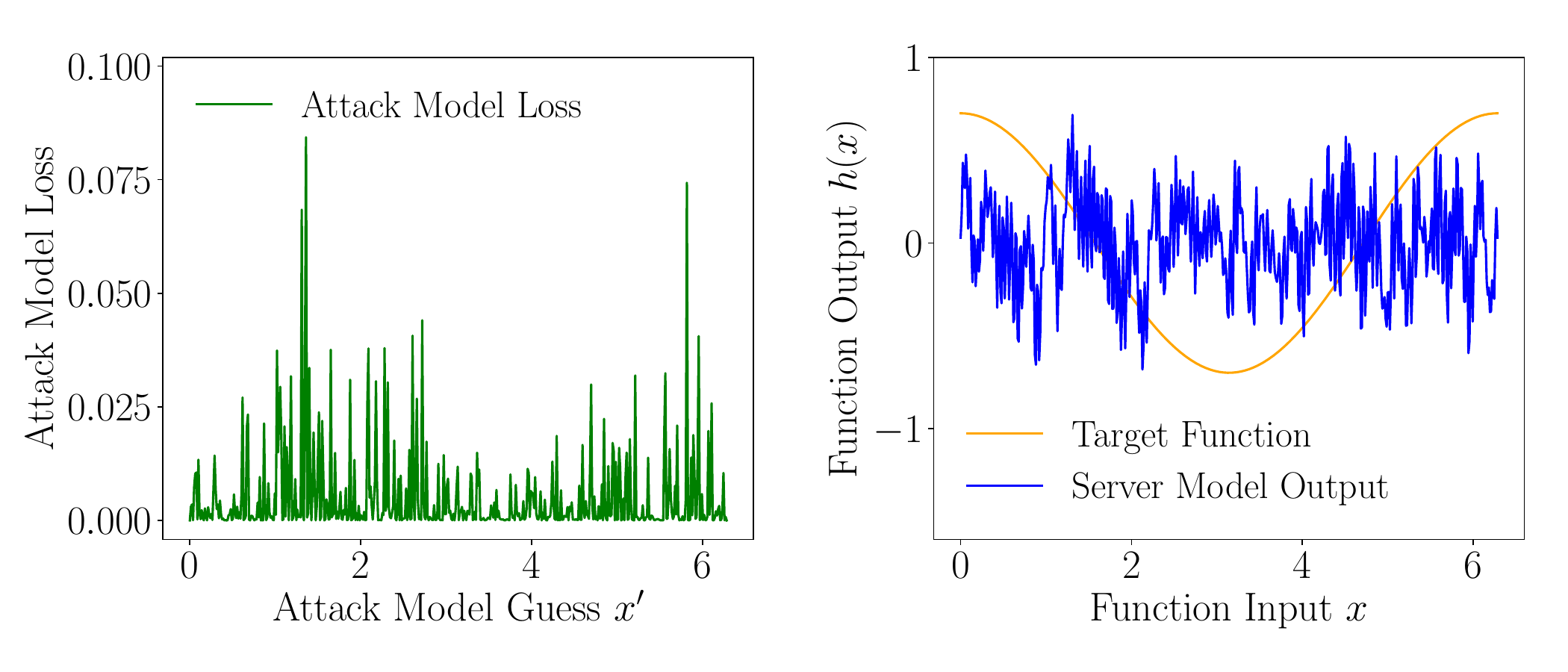}
        \caption{Attack loss and FL model fit at epoch $t = 0$}
    \end{subfigure}
    \begin{subfigure}{0.48\textwidth}
        \includegraphics[width=\textwidth]{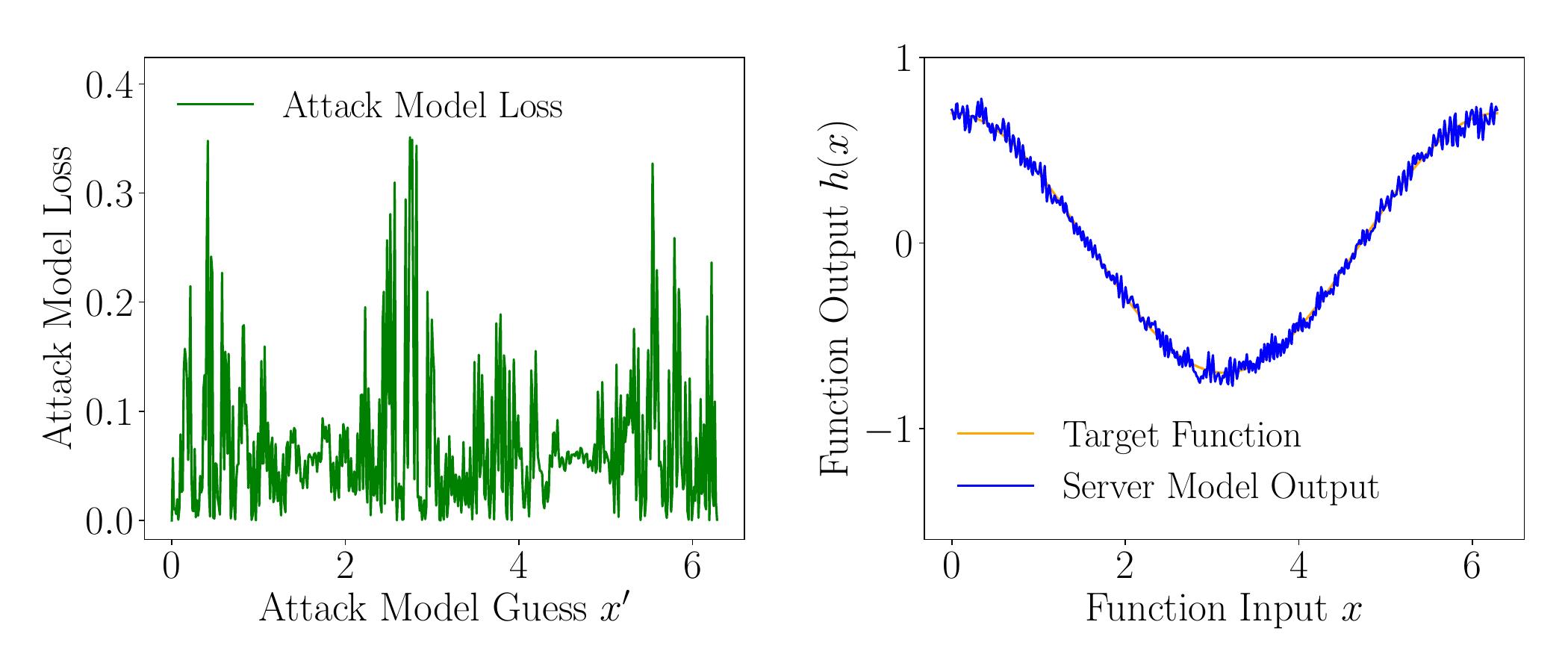}
        \caption{Attack loss and FL model fit at epoch $t = 100$}
    \end{subfigure}
    \begin{subfigure}{0.48\textwidth}
        \includegraphics[width=\textwidth]{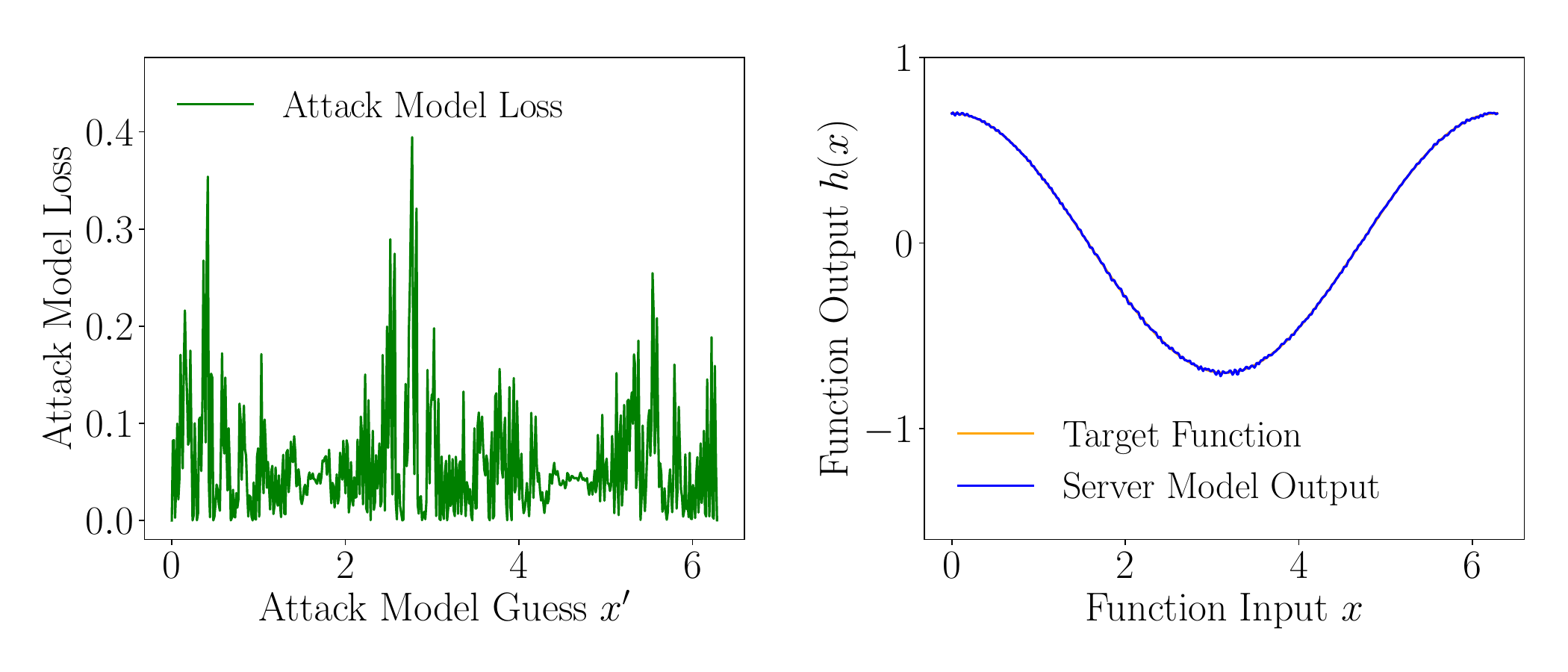}
        \caption{Attack loss and FL model fit at epoch $t = 400$}
    \end{subfigure}   
    \caption{(Left) The loss landscape when matching gradients during a gradient based attack at various points during the FL model's training. (Right) FL model output as it is trained to fit a simple cosine function $h(x)=0.7\cos {x}$. The loss landscape was calculated using the current thetas of the FL model, with the loss calculated using the gradients generated by $x=1.6371$ and $y=h(x)$. The model was univariate with a single $x$ input and four qubits per data point $m=4$.}
    \label{fig:landscape_flattening_solution}
\end{figure}

We demonstrate an adversarial example of an underlying distribution that could reduce the privacy of the quantum federated learning model in Fig.~\ref{fig:landscape_flattening}. When using a relatively low-frequency model $m=2$, the FL model is able to match the target function of $y=0.7 \cos{x}$ almost exactly, and in the process suppresses high-frequency terms. This leads to a simplified gradient attack model loss landscape with reduced privacy. However, we further show in Fig.~\ref{fig:landscape_flattening_solution} that a higher frequency model $m=4$ trained on the same function $y=0.7 \cos{x}$ ends up achieving a good fit to the target function, but maintains the high-frequency terms and thus the privacy. This property of quantum models preserving high-frequency terms has been shown to be useful when data is drawn from some signal distribution with an added error, as the high-frequency terms allow overfitting to the data, while still providing good generalization to the signal function overall \cite{peters2022generalization}. In our context the preservation of high-frequency terms during training means that the privacy is maintained while the FL model can still achieve good generalization for low frequency functions.  

\section{Conclusion} \label{sec:conclusion}

In this work, we highlight that variational quantum circuits consisting of expressive encoding maps and overparameterized ans\"atze naturally provide privacy in the context of federated learning. We discern that expressive quantum circuits inherently possess a feature map which transforms classical data into a higher-dimensional Hilbert space resulting in quantum models that encompass high-degree Fourier frequencies. This results in the necessity to solve a system of very high degree Chebyshev polynomials in the input space in order to learn the underlying data. We analyze various techniques of solving such a system of equations, either via trying to solve them analytically, or by using a machine learning-based attack. Our observation is that all these methods lead to an exponential (in the number of qubits) hardness in recovering the underlying client's input. 

We primarily notice that expressive quantum circuits, when underparameterized in terms of optimizable  variables, give rise to hard to train loss landscapes in the machine learning-based attack. In the case studied here, the attack model's underparameterization in the optimizable variable $\mathbf{x}'$ renders it challenging to train, while, conversely, the FL model being overparameterized in its optimizable variable $\boldsymbol \theta$ ensures it can be trained without suffering from the issue of spurious local minima. 

This inherent dichotomy between overparameterization and underparameterization ensures that while FL models can excel in their designated learning tasks, they simultaneously offer enhanced privacy. Both our theoretical and empirical findings indicate that high-frequency terms contribute to periodic loss functions that are exponentially hard to optimize. This revelation prompts the intriguing question: \emph{Could classical federated learning techniques potentially benefit from incorporating highly periodic feature maps during a data preprocessing phase, drawing inspiration from quantum circuit feature maps?} 

It remains an open question whether classical machine learning methods can effectively navigate this feature space, or if expressive variational quantum circuits are intrinsically better suited for training within these highly periodic feature domains. Future research could also explore the efficacy of quantum encodings which are hard to simulate classically. In  this context, our work introduces a novel paradigm in quantum machine learning where a quantum algorithm's success isn't solely gauged by its ability to outperform its classical counterpart in terms of model metrics, but also by its potential to offer superior privacy.

\section*{Author Contribution}

N. Kumar devised the project with contribution from S. Eloul based on his previous work \cite{eloul2022enhancing}. N. Kumar, C. Li and J. Heredge developed the connection of expressivity of quantum models in federated learning. J. Heredge and N. Kumar developed the privacy analysis. J. Heredge performed the numerical simulations showcasing privacy. C. Li and S.H. Sureshbabu made the figures illustrating federated learning setup. M. Pistoia led the overall project. All authors contributed in writing the manuscript. 

\section*{Acknowledgments}
The authors thank Shouvanik Chakrabarti, Dylan Herman, Enrico Fontana, and the other colleagues at the Global Technology Applied Research Center of JPMorgan Chase for support and helpful discussions.

\bibliographystyle{ieeetr}
\bibliography{bibliography}
\section*{Disclaimer}

This paper was prepared for informational purposes by the Global Technology Applied Research center of JPMorgan Chase $\&$ Co. This paper is not a product of the Research Department of JPMorgan Chase $\&$ Co. or its affiliates. Neither JPMorgan Chase $\&$ Co. nor any of its affiliates makes any explicit or implied representation or warranty and none of them accept any liability in connection with this paper, including, without limitation, with respect to the completeness, accuracy, or reliability of the information contained herein and the potential legal, compliance, tax, or accounting effects thereof. This document is not intended as investment research or investment advice, or as a recommendation, offer, or solicitation for the purchase or sale of any security, financial instrument, financial product or service, or to be used in any way for evaluating the merits of participating in any transaction.

\end{document}